\documentclass{ws-ijmpa} 
\usepackage{hyperref}
\usepackage{epsfig,rotating}
\usepackage{amsmath,amssymb}
\usepackage{dsfont}

\newcommand{\be}{\begin{equation}}
\newcommand{\ee}{\end{equation}}
\newcommand{\bea}{\begin{eqnarray}}
\newcommand{\eea}{\end{eqnarray}}
\newcommand{\nn}{\nonumber}
\newcommand{\vx}{\vec{x}}
\newcommand{\vX}{\vec{X}}
\newcommand{\vp}{\vec{p}}
\newcommand{\vP}{\vec{P}}
\newcommand{\vq}{\vec{q}}
\newcommand{\vQ}{\vec{Q}}
\newcommand{\vk}{\vec{k}}
\newcommand{\vK}{\vec{K}}
\newcommand{\oO}{\overline{\Omega}}

\begin{document}

\markboth{Jun Wu, Jimmy A. Hutasoit, Daniel Boyanovsky, Richard Holman}
{Neutrino Oscillations, Entanglement and Coherence}


\catchline{}{}{}{}{}


\title{NEUTRINO OSCILLATIONS, ENTANGLEMENT AND COHERENCE:\\ A Quantum Field Theory Study In Real Time}

\author{Jun Wu}

\address{Department of Physics and
Astronomy, University of Pittsburgh, \\
Pittsburgh, PA 15260 \\
juw31@pitt.edu}

\author{Jimmy A. Hutasoit}

\address{Department of Physics, Carnegie Mellon University, \\
Pittsburgh, PA 15213, USA\\
jhutasoi@andrew.cmu.edu}

\author{Daniel Boyanovsky}

\address{Department of Physics and Astronomy, University of Pittsburgh, \\
Pittsburgh, PA 15260\\
boyan@pitt.edu}

\author{Richard Holman}

\address{Department of Physics, Carnegie Mellon University, \\
Pittsburgh, PA 15213, USA\\
rh4a@andrew.cmu.edu}

\maketitle

\begin{history}
\received{\today}
\revised{Day Month Year}
\end{history}


\begin{abstract}
The dynamics of neutrino mixing and oscillations are studied directly in finite \emph{real time} in a model that effectively describes charged current weak interactions. Finite time corrections to the S-matrix result  for the appearance and disappearance probabilities are obtained. It is observed that these effects may be of the same order of the S-matrix result in long-baseline appearance experiments. We argue that fundamentally, the S-matrix is ill-suited
to describe long-baseline events due to the fact that the neutrino is produced in an entangled state with the charged lepton, which can be disentangled by the measurement of the charged lepton near the production site. The appearance and disappearance far-detection process is described from the time evolution of this disentangled ``collapsed'' state, allowing
us to establish the conditions under which factorization of detection rates emerges in long-baseline experiments. We also study the time evolution of the reduced density matrix and
show explicitly how oscillations are manifest in the off-diagonal terms, {\it i.e.}, coherences, as a result of a finite time analysis. Lastly, we study a model for the ``GSI  anomaly'' obtaining the time evolution of the population of parent and daughter particles directly in real time. We confirm that the decay rate of parent and growth rate of
daughters do \emph{not} feature oscillatory behavior from interference of mass eigenstates.

\keywords{neutrinos oscillation; quantum field theory; entanglement.}
\end{abstract}

\ccode{PACS numbers: 14.60.Pq;13.15.+g;12.15.Ff}


\section{Introduction}
Neutrino masses, mixing and oscillations are the clearest evidence yet of physics beyond the standard model \cite{book1,book2,book3}. They provide an explanation of the solar neutrino problem \cite{msw,book4,haxtonsolar1,haxtonsolar2} and have other important phenomenological \cite{book1,book2,book3,grimuslec,kayserlec1,kayserlec2,mohapatra1,mohapatra2,degouvea1,degouvea2,degouvea3},
astrophysical \cite{book4,book5,haxton} and cosmological \cite{dolgovcosmo1,dolgovcosmo2} consequences. Moreover, another fascinating aspect of mixing and oscillations of ultrarelativistic neutrinos is that they provide a remarkable manifestation of \emph{macroscopic quantum coherence} over
unprecedented long distance and time scales. Whereas in condensed matter systems macroscopic quantum coherence is maintained over mesoscopic scales, long baseline disappearance and appearance experiments probe the coherence of the neutrino states over scales of hundreds of kilometers. \\

In its simplest (and perhaps overly naive) inception, mixing and oscillations of massive neutrinos have been explained by an analogy with a two level system undergoing Rabi-like oscillations between them. Oscillations emerge from the interference of the quantum mechanical states associated with the mass eigenstates (see for example \cite{book1,book2,book3,kayserlec1,kayserlec2,grimuslec} and references therein). As simple and compelling  as this interpretation is, deeper investigations of this basic paradigm have raised a number of important and fundamental questions \cite{aksmir1}. One of these involves the energy and momentum uncertainties \cite{kayser1,kayser2,rich,nauenberg,lipkin1,lipkin2,lipkin3}, a topic that is still receiving attention \cite{lipkinuncertainty1,lipkinuncertainty2,lipkingsi,glashow} (for a recent review see \cite{akmerev}). The recognition that exact energy and momentum conservation prevent oscillations between
neutrinos of different masses \cite{kayser1,kayser2} has led to the consideration of oscillation experiments in terms of wave-packets \cite{kayser1,kayser2,rich,nauenberg} including the quantum
mechanical aspects of production and detection which have been incorporated in the quantum field theoretical framework \cite{kiers1,kiers2,grimus1,grimus2,grimus3,grimus4,grimus5,giunti1,giunti2,giunti3,giunti4,giunti5,giunticohe,pila,cardall,beuthe,dolgov1,dolgov2,dolgov3,boyaho,keister,akmerev}. \\

Previous quantum field theory treatments of mixing and oscillations are S-matrix theoretic in nature, making use of  in-out wave-packets spatially localized at the source, or the ``near" detector\footnote{We are using the term ``near" detector to describe measurement apparatus at or nearby the production site, which is used to detect the charged lepton that is produced with the primary neutrino. This is not to be confused with the term near detector that is widely used in the experimental literatures (see for example \cite{Ellis:2006wj}). The latter is used to detect neutrino at a short baseline.}, and the far detector \cite{grimus1,grimus2,grimus3,grimus4,grimus5,giunti1,giunti2,giunti3,giunti4,giunti5,giunticohe,pila,cardall,beuthe,dolgov1,dolgov2,dolgov3,keister,akmerev}.
However, in all  of these treatments the S-matrix calculation takes the \emph{interaction time to infinity}, even when in some treatments the initial and final wave packets are defined at finite source and detector times \cite{cardall,dolgov1,dolgov2,dolgov3,keister}.  In the wave-packet-S-matrix approach advocated in many of the references above, the wave packets spatially localized at the source   and far detectors   respectively  are spatially separated by the finite baseline, and \emph{also} localized in time, however, the intermediate steps invoke the S-matrix approach wherein in and out states are prepared at $t_i=-\infty ,t_f=\infty$  .

This incongruity between keeping a finite baseline, with wave packets defined at some initial time at the source and final time at the far detector, and taking the time to infinity in the S-matrix element is
usually justified with the statement that time is not measured in appearance or disappearance experiments. \\

While this is true, it is a practical, but not a fundamental reason for taking the infinite time limit. One can consider a \emph{gedanken} experiment in which clocks at the source (near detector) and far detector are \emph{synchronized} via global positioning satellites
and register the detection of the charged leptons at the source and far detectors at the time $t_S$ and $t_D$, respectively. Obviously, the time difference registered by these clocks $t_D-t_S \sim L$, with $L$ being the baseline. This has uncertainties of the order of the size of the source and detectors as the interaction vertices from which the charged leptons emerge
are localized within these regions (although current  resolution of the interaction vertices is much more accurate than this scale). \\

Furthermore, taking the infinite time limit as in the usual  S-matrix calculation enforces total
energy conservation via an overall delta function in the transition amplitude. The transition probability treats the square of this delta function as overall energy conservation multiplied by the total time of the interaction, from which an \emph{interaction rate} is extracted by dividing by this time (in the long time limit). However, oscillations arising from quantum mechanical interference have nothing to do with a transition \emph{rate} and in principle do not feature a secular evolution in time. \\

While there are various quantum field theory calculations of appearance and disappearance processes in the literature (see references above), all of them invoke the S-matrix approach and take the infinite time limit from the outset. It is fundamentally important to understand any possible caveat in S-matrix approach by focusing on the fact that only a finite time interval elapses between the production and the detection of neutrinos.
In this article, we aim to demonstrate that \emph{conventional} S-matrix calculation where in and out states are eigenstates of energy-momentum is ill-suited in principle to describe neutrino oscillations by studying the quantum field theory of neutrino mixing and oscillations in \emph{real time} in the plane wave limit. We keep a finite time interval between the initial state ``prepared'' at the source and the final state measured at the (far) detector. After obtaining the transition amplitudes and probabilities, we point out the problems in the standard S-matrix calculation. Then, combined with our recent work on the disentanglement issue associated with neutrino oscillations \cite{NuDynamics10}, we seek a complete space-time description of the dynamics of mixing and oscillations that describes long-baseline experiments in the plane wave limit. \\


We obtain the transition matrix element in which neutrinos are produced and detected via charged ``leptons'', directly in real time. Comparing with the S-matrix results in the plane wave basis, we discuss how wave packet localization restricts the contributions from different channels and obtain the non-conventional transition probabilities at finite but long time. For the \emph{appearance probability}, we find that finite time corrections are comparable to or larger than the S-matrix contribution. \\

Then we argue that this formulation is incorrect because long-baseline experiments involve detection measurements at \emph{two different} times, corresponding to detection or ``collapse" of the charged leptons at the near and the far detectors.
The neutrino is produced in an entangled state at the interaction vertex at the source, but the measurement of the charged lepton at the near detector \emph{disentangles} the neutrino. It is this ``collapsed'' disentangled state that evolves in time and leads
to the production of the charged lepton which is measured in the far detector \cite{giunticohe,glashow,losecco1,losecco2}. We carry out a systematical study containing the dynamics of entanglement and disentanglement, namely that neutrinos produced by the decay of a parent particle are \emph{entangled} with the daughter charged lepton and then disentangled close to the near detector. Different from the treatment \cite{giunticohe,glashow} invoking the infinite time limit, we analyze the time evolution of the reduced density matrix for the neutrino and show that keeping finite time allows us to understand the time evolution of the population and coherences. We establish that indeed coherence, as determined by the off-diagonal density matrix elements in the mass basis is maintained up to the oscillation time scale \cite{NuDynamics10}. \\

We then obtain the near and far detector event rates from the detailed evolution of the entangled state, disentanglement at the near detector and further evolution to the far detector. We discuss under which circumstances the factorization of the processes is
valid. We also discuss in detail under what circumstances the usual quantum mechanical description is valid. The real-time analysis of disentanglement and coherence shows that
this is the case provided the neutrino state is disentangled on time scales much shorter than the oscillation scale. This analysis also makes manifest the compatibility with energy
conservation. \\

Another clear evidence of the fact that the theory  of neutrino oscillations is far from being understood at a fundamental level is found in recent controversies regarding the recoilless emission and detection of neutrinos (Mossbauer neutrinos) \cite{rag1,rag2,bilenky,akhme1,akhme2}. Yet another can be found in the controversial interpretation of the GSI anomaly \cite{giuntygsi1,giuntygsi2,lipkingsi,burka,kienert,peshkin,Alexander:2009}, namely, periodic modulations in the K-electron capture and $\beta^+$ decay rate \cite{gsi1,gsi2} in terms of quantum beats resulting from the mixing of neutrino mass eigenstates and their interference in the final state \cite{kienle,ivanov1,ivanov2,ivanov3,faber,Alexander:2009}.
In this article, we review our real time analysis of the decay of a parent particle into a daughter particle and a ``flavor'' neutrino \cite{GSI-wu}, as an application of the real time method developed for neutrino oscillations.
We show that both the decay rate of the parent, and growth rate of the daughter populations \emph{do not feature oscillations} as a consequence of mixing and establish unambiguously the reasons. We therefore conclude that the ``GSI-anomaly'' cannot be explained by the interference of mass eigenstates in the final state.

\section{A   Model of ``Neutrino'' Oscillations}\label{sec:model}
The goal of this work is to study the dynamics of mixing and oscillation of neutrinos in quantum field theory but in a finite time interval. In order to exhibit the main results in a clear and simpler manner, we introduce a bosonic model that describes mixing, oscillations and charged current weak interactions reliably without the complications associated with fermionic and gauge fields. We can do so because the technical complications associated with spinors and gauge fields are irrelevant to the physics of mixing and oscillations, as is obviously manifest in meson mixing. Our model is defined by the following Lagrangian density
\be
\mathcal{L} =
\mathcal{L}_0[W,l_\alpha]+ \mathcal{L}_0[\nu_\alpha] +
\mathcal{L}_{\rm int}[W,\l_{\alpha},\nu_\alpha] ~~;~~ \alpha=e,\mu
\label{totallag} \ee with \be {\cal L}_0[\nu] =
\frac{1}{2}\left[\partial_{\mu}\Psi^T
\partial^{\mu}\Psi  -\Psi^{T} \mathbb{M}\Psi  \right] \label{nulag}\, ,
\ee
where  $\Psi$ is a flavor doublet representing the neutrinos
\be
\Psi = \left(
             \begin{array}{c}
               \nu_e \\
               \nu_\mu \\
             \end{array}
           \right), \label{doublet}\ee and $\mathbb{M}$ is the mass matrix
            \be \mathbb{M} = \left(
                               \begin{array}{cc}
                                 m_{ee} & m_{e\mu} \\
                                 m_{e\mu} & m_{\mu \mu} \\
                               \end{array}
                             \right)\,. \label{massmtx}
\ee
The interaction Lagrangian is similar to the charged current interaction of the standard model, namely
\be
{\cal L}_{\rm int} (\vx,t) = g\, W(\vx,t)\Big[ l_{e}( \vec{x},t)\,\nu_{e}(
\vec{x},t)+ l_{\mu}( \vec{x},t)\,\nu_{\mu}( \vec{x},t)\Big],
\label{Interaction}
\ee
where $g$ is the coupling constant. $W(x)$ represents the vector boson, or alternatively the pion field, and $l_{\alpha}$, $\alpha = e,\mu$ the two charged leptons.
From the interaction Lagrangian (\ref{Interaction}), it is clear that $W$, $l_{\alpha}$ and $\nu_{\alpha}$ are all hermitian.
The mass matrix is diagonalized by a unitary transformation
\be
U^{-1}(\theta) \, \mathbb{M}\, U(\theta) = \left(
                                                           \begin{array}{cc}
                                                             m_1 & 0 \\
                                                             0 & m_2 \\\end{array}\right)
 ~~;~~ U(\theta) = \left(                                                                                  \begin{array}{cc}                                                                                       \cos \theta & \sin \theta \\                                                                                       -\sin\theta & \cos\theta \\                                                                                     \end{array}                                                                                   \right) . \label{massU}
\ee
In terms of the doublet of mass eigenstates,  the flavor doublet can be expressed as
\be
\left(
                                     \begin{array}{c}
                                       \nu_e \\
                                       \nu_\mu \\
                                     \end{array}
                                   \right) = U(\theta)\,\left(
                                               \begin{array}{c}
                                                 \nu_1 \\
                                                 \nu_2 \\
                                               \end{array}
                                             \right) \,.\label{masseigen}
\ee
This bosonic model clearly describes charged current weak
interactions reliably as it includes all the relevant aspects of
mixing and oscillations. \\

\begin{figure}[h]
\begin{center}
\includegraphics[width=6cm,keepaspectratio=true]{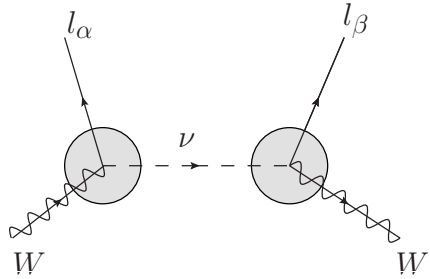}
\caption{Typical experiment in which the charged leptons are measured at a near and far
detector and the neutrino is an intermediate state.}
\label{fig:experiment setup}
\end{center}
\end{figure}

We consider ``neutrino'' oscillation experiments following the interaction processes illustrated in Fig.~\ref{fig:experiment setup},
\bea
W\rightarrow l_{\alpha}+\nu_{\alpha} \rightsquigarrow & \left\{\begin{array}{ll} W+l_{\beta}~~,\beta \neq \alpha & \mbox{
appearance  } \\ W+l_{\alpha} & \mbox{ disappearance
 } \end{array}\right.\,.
\eea

In a quantum field theory calculation of the transition amplitudes, neutrinos propagating between the production  and detection regions are considered as intermediate particles
\cite{rich,kiers1,kiers2,grimus1,grimus2,grimus3,grimus4,grimus5,giunti1,giunti2,giunti3,giunti4,giunti5,giunticohe,dolgov1,dolgov2,dolgov3,cardall,pila,keister,akmerev} and are described by a (free field) propagator. \\

In order  to extract the main physical aspects and new results in the clearest manner, we study the time evolution in terms of  \emph{plane waves}, postponing the necessarily more technical  discussion in terms of  wave-packets to a follow up article. However, to set the stage to the next step that includes such formulation, we revisit some aspects related to wave packet localization that are relevant for the present discussion. \\

For a wave packet description we consider the initial and final $W,l_{e,\mu}$ states to be described by wave packets. Specifically, we consider the following type of spatially localized states
\be
f(\vx,\vX;\vp) = N~e^{-\frac{(\vx-\vX)^2}{2\sigma^2}}~e^{-i\vp\cdot\vx}
\label{WPx}
\ee
where $N$ is a normalization factor, $\vX_{S,D}$ is the center of the wave packet for source and detection, respectively. $\sigma$ is the localization length of the wave packet, which might be a nuclear scale or a \emph{macroscopic} scale of the order of the size of the detector and production regions. We will only require that  $\sigma \ll |\vX_{D}-\vX_S|$. The spatial
Fourier transform of these wave packets is
\be
\tilde{f}(\vP,\vp;\vX)
= \int d^3x e^{ i\vP\cdot\vx} f(\vx,\vX;\vp) = \mathcal{N}~e^{-
\frac{1}{2}(\vP-\vp)^2\sigma^2}~e^{
i(\vP-\vp)\cdot\vX}~~;~~\mathcal{N} = \Bigg[
\frac{\sigma}{\sqrt{2\pi}}\Bigg]^3. \label{FTf}
\ee
The normalization factor in Fourier space, $\mathcal{N}$, has been chosen so that in the
macroscopic limit $\sigma \rightarrow \infty$, the wave packet has a definite momentum
$\tilde{f}(\vp,\vk;\vX)\rightarrow \delta^{(3)}(\vp-\vk)$. The initial and final quantum states are localized at the source ($\vX_S$) or detection ($\vX_D$) sites and are described by wave packets that feature mean momenta $\vK_S;\vK_D;\vP_S,\vP_D$ respectively

\bea |\widetilde{W}(S)\rangle \equiv|\widetilde{W}(\vK_S;\vX_S)\rangle & = & \int d\vk_S~
\tilde{f}(\vK_S,\vk_S;\vX_S)~|W(\vk_S)\rangle, \nonumber \\
 |\widetilde{W}(D)\rangle \equiv |\widetilde{W}(\vK_D;\vX_D)\rangle & = &
 \int d\vk_D~\tilde{f}(\vK_D,\vk_D;\vX_D)~|W(\vk_D)\rangle, \nonumber \\
 |\widetilde{l}_\alpha(S)\rangle \equiv|\widetilde{l}_\alpha (\vP_S;\vX_S)\rangle
  & = & \int d\vp_S~\tilde{f}(\vP_S,\vp_S;\vX_S)~|l_\alpha (\vp_S)\rangle, \nonumber
   \\ |\widetilde{l}_\alpha (D)\rangle \equiv |\widetilde{l}_\alpha (\vP_D;\vX_D)\rangle
   & = & \int d\vp_D~\tilde{f}(\vP_D,\vp_D;\vX_D)~|l_\alpha (\vp_D)\rangle, \label{locstates}
\eea
where the quantum states $|W(\vk_S)\rangle\;, |W(\vp_D)\rangle\,,|l_\alpha (\vk_S)\rangle\,,|l_\alpha (\vp_D)\rangle$ on the right hand side of Eq.~(\ref{locstates}) are \emph{plane wave} single particle states, which are the eigenstates of the non-interacting Hamiltonian. In term of these localized states, the appearance and disappearance transition amplitudes \emph{for wave packets} prepared and detected at the source (near detector) at an initial time $t_i$ and detected at a far detector at a final time $t_f$ are given by
\bea
\widetilde{\mathcal{A}}_{\alpha\rightarrow\beta} & =&  \langle \widetilde{W} (D), \widetilde{l}_{\beta}(D); \widetilde{l}_{\alpha}(S) |e^{-i\hat{H}(t_f-t_i)}|\widetilde{W} (S)\rangle,   \label{amplitudeab}\\
\widetilde{\mathcal{A}}_{\alpha\rightarrow\alpha} & = &  \langle
\widetilde{W} (D), \widetilde{l}_{\alpha}(D);
\widetilde{l}_{\alpha}( S) |e^{-i\hat{H}(t_f-t_i)}|\widetilde{W}(S)
\rangle, \label{amplitudeaa} \eea
where $ D$ and $ S$ label the states localized in the detection and source regions,
respectively. The respective probabilities are
\be
\mathcal{P}_{\alpha\rightarrow\beta}    =
|\widetilde{\mathcal{A}}_{\alpha\rightarrow\beta}|^2 \qquad {\rm and} \qquad  \mathcal{P}_{\alpha\rightarrow\alpha}    =
|\widetilde{\mathcal{A}}_{\alpha\rightarrow\alpha}|^2 \,.
\label{probabilityab}
\ee

The transition amplitudes between the initial and final localized states are given by
\bea
& & \langle \widetilde{W} (D), \widetilde{l}_{\beta}(D); \widetilde{l}_{\alpha}(S) |e^{-i\hat{H}(t_f-t_i)}|\widetilde{W} (S)\rangle \nn \\
&=& \int d\vk_S~ d\vp_S~ d\vp_D~ d\vk_D \,\, e^{-iE_f t_f} ~\mathcal{A}_{\alpha \rightarrow \beta}~ e^{iE_i t_i} \, \tilde{f}^*(\vK_D,\vk_D;\vX_D) \nonumber \\ &&
\tilde{f}^*(\vP_S,\vp_S;\vX_S)\,\tilde{f}^*(\vP_D,\vp_D;\vX_D)\, \tilde{f}(\vK_S,\vk_S;\vX_S), \label{amploc}
\eea
where
\be
E_i = E^W_{\vk_S}~~;~~E_f=E^W_{\vk_D}+E^l_{\vp_D}+E^l_{\vp_S}, \label{inifinener}
\ee
and $t_i$ and $t_f$ label the initial and final times.
\be
\mathcal{A}_{\alpha \rightarrow \beta} = \langle {W} (\vk_D), {l}_{\beta}(\vp_D); {l}_{\alpha}(\vp_S) |U(t_f,t_i)| {W} (\vk_S)\rangle \label{Mfi}
\ee
is the usual transition matrix element in terms of plane waves eigenstates of the non-interacting Hamiltonian in the interaction picture. To obtain the above expressions we have passed to the interaction picture by writing
\be
e^{-i\hat{H}(t_f-t_i)} = e^{-iH_0 t_f}~U(t_f,t_i)~ e^{ iH_0 t_i} \,,\label{intpic}
\ee
with the time evolution operator in the interaction picture
\be
U(t_f,t_i) = e^{ iH_0 t_f}~e^{-i\hat{H}(t_f-t_i)}~e^{- iH_0 t_i} =
 T~ e^{i \int_{t_i}^{t_f}d^3x dt ~\mathcal{L}_{\rm int}(\vx,t)}\,, \label{timeevolop}
\ee
and we have also used the fact that the initial and final quantum states in (\ref{Mfi}) are free single particle plane wave eigenstates of the non-interacting Hamiltonian $H_0$ with $E^W_{\vk}=\sqrt{k^2+M^2_W}$, $E^l_{\vp}=\sqrt{p^2+m^2_l}$. \\

These are the main ingredients in the appearance and disappearance transition probabilities in terms of the localized wave packets. In the limit when the localization length goes to infinity $\sigma \rightarrow \infty$ with the normalization (\ref{FTf}), the functions $\tilde{f}$ become delta functions in momentum and the localized states become simple plane wave states.
In what follows we will obtain the transition matrix element $\mathcal{A}_{\alpha \rightarrow\beta}$ (\ref{Mfi}) up to second order in the interaction.

\section{Appearance and disappearance amplitudes and probabilities}\label{sec:3}

\subsection{Localization and suppression of crossed channel contributions}
\begin{figure}[h]
\begin{center}
\includegraphics[width=12cm,keepaspectratio=true]{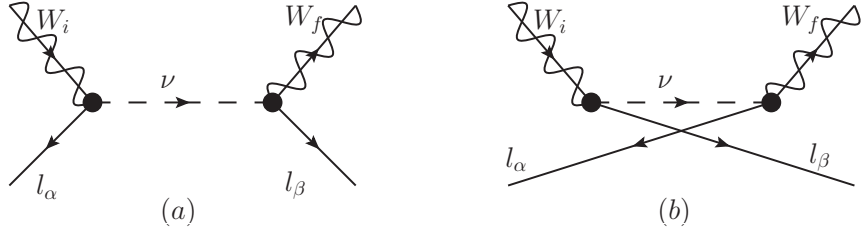}
\caption{Feynman diagrams that contribute to
$\mathcal{A}_{\alpha\rightarrow \beta}$ up to second order.
 Diagram (a) corresponds to the ``s-channel'' and (b) to the ``t-channel." }
\label{fig:FeynmanDiagram}
\end{center}
\end{figure}

Up to second order in the coupling $g$ the matrix element
$\mathcal{A}_{\alpha\rightarrow \beta}$ features the  contributions
depicted in figs. (\ref{fig:FeynmanDiagram}a,b) respectively. If the
lepton at the production vertex were in-coming, Fig. (a) would be
the equivalent of an s-channel and Fig. (b) of a t-channel contribution.
We refer to these as ``s'' and ``t'' channel contributions
respectively for the remainder of the discussion.

For the ``s''  and ``t''-channels we find respectively

\bea
\mathcal{A}^{(s)}_{\alpha \rightarrow \beta} = -g^2~\Pi~\int_{t_i}^{t_f}dt_1dt_2\int d^3x_1 d^3x_2&&e^{i(E^{W}_{\vk_D}t_1-\vec{k}_D\cdot\vec{x}_1)}\,e^{i(E^{l}_{\vp_D}t_1-\vec{p}_D\cdot\vec{x}_1)}\, e^{i(E^{l}_{\vp_S}t_2-\vec{p}_S\cdot\vec{x}_2)}\nonumber \\
&&e^{-i(E^{W}_{\vk_S}t_2-\vec{k}_S\cdot\vec{x}_2)} \,\,\langle 0 |T\big(\nu_{\beta}(x_1)\nu_{\alpha}(x_2)\big)|0\rangle \,,
\label{Mschannel}
\eea
and
\bea \mathcal{A}^{(t)}_{\alpha \rightarrow \beta} =
-g^2~\Pi~\int_{t_i}^{t_f}dt_1dt_2\int d^3x_1 d^3x_2&&e^{i(E^{W}_{\vk_D}t_1-\vec{k}_D\cdot\vec{x}_1)}\,e^{i(E^{l}_{\vp_S}t_1-\vec{p}_S\cdot\vec{x}_1)}\,
e^{i(E^{l}_{\vp_D}t_2-\vec{p}_D\cdot\vec{x}_2)} \nonumber \\
&&e^{-i(E^{W}_{\vk_S} t_2-\vec{k}_S\cdot\vec{x}_2)}\,\, \langle 0
|T\big(\nu_{\beta}(x_1)\nu_{\alpha}(x_2)\big)|0\rangle \,,
\label{Mtchannel}
\eea
where for disappearance $\alpha = \beta$, for appearance $\alpha \neq \beta$, and
\be
\Pi = \Big[\frac{1}{16~V^4~E^W_{\vk_S}\,E^W_{\vk_D}\,E^l_{\vp_S}\,E^l_{\vp_D}}
\Big]^\frac{1}{2} ~~\,. \label{Pifactor}
\ee

The transition matrix elements $\widetilde{\mathcal{A}}_{\alpha\rightarrow \beta}$ between the initial and final localized states are obtained from the $\mathcal{A}_{\alpha \rightarrow \beta}$ matrix elements above by convolution with the wave packets describing these localized states (see Eq. \ref{amploc}). In order to understand the features associated with the localization, we consider wave packets with a \emph{macroscopic} localization scale
$\sigma \ll |\vec{X}_D-\vec{X}_S|$ of the order of the size of the detectors, in which case the wave packets are nearly plane waves and we can approximate $ p_{S,D} \simeq  P_{S,D}$, $k_{S,D} \simeq  K_{S,D}$ in the arguments. For the s-channel we obtain
\bea
\widetilde{\mathcal{A}}^{(s)}_{\alpha \rightarrow \beta}   \propto e^{-\frac{(\vx_1-\vX_S)^2}{\sigma^2}}~e^{-\frac{(\vx_2-\vX_D)^2}{\sigma^2}} \,,\label{schannover}
\eea
whereas for the t-channel contribution we
obtain
\bea \widetilde{\mathcal{A}}^{(t)}_{\alpha \rightarrow \beta}   \propto
e^{-\frac{(\vx_1-\vX_S)^2}{2\sigma^2}}~e^{-\frac{(\vx_1-\vX_D)^2}{2\sigma^2}}~
e^{-\frac{(\vx_2-\vX_S)^2}{2\sigma^2}}~e^{-\frac{(\vx_2-\vX_D)^2}{2\sigma^2}}\,.\label{tchannover}
\eea
Therefore, while for the s-channel the $x_1,x_2$ integrals factor out, the t-channel contribution obviously vanishes for $|\vX_D-\vX_S|\gg \sigma$. In other words, a localization    of the initial and final states even over \emph{macroscopic} scales, as long as  $\sigma \ll |\vec{X}_D-\vec{X}_S|$, ensures that the ``t''-channel contribution depicted in the second
Feynman diagram Fig.~\ref{fig:FeynmanDiagram} vanishes.

\subsection{Amplitudes and probabilities for plane waves}\label{sec:probs}

In what follows we consider only the ``s''-channel contribution and
focus solely on studying the transition amplitudes for \emph{plane
waves}, postponing to forthcoming work a full space-time description
of the production-detection process. \\

The plane wave transition amplitudes  $ \mathcal {A}_{\alpha \rightarrow \beta}$ are given by \bea
\mathcal {A}_{\alpha \rightarrow \beta} = -g^2~\Pi~\int_{t_i}^{t_f}dt_1dt_2\int d^3x_1
d^3x_2&& e^{i(E_D^{W}t_1-\vec{k}_D\cdot\vec{x}_1)}\, e^{i(E_{D}^{l}t_1-\vec{p}_D\cdot\vec{x}_1)}\,e^{i(E_S^{l}t_2-\vec{p}_S\cdot\vec{x}_2)}\nonumber \\
&&e^{-i(E_S^{W}t_2-\vec{k}_S\cdot\vec{x}_2)}\,\,\langle 0|T\big(\nu_{\beta}(x_1)\nu_{\alpha}(x_2)\big)|0\rangle \,.
\label{Mapp-1}
\eea
Here
\be
\langle 0|T\big( \nu_{\beta}(x_1)\nu_{\alpha}(x_2)\big)|0\rangle = \sum_{j=1,2}U_{\beta j} U_{j \alpha} \langle 0|T\big( \nu_{j}(x_1)\nu_{j}(x_2)\big)|0\rangle \label{flavorpropa}\,,
\ee
where $U_{ab}$ are the matrix elements of the matrix $U$ given in Eq. (\ref{massU}),  and
\be
\langle 0|T\big( \nu_{j}(x_1)\nu_{j}(x_2)\big)|0\rangle =  i ~\int\frac{d^3p}{(2\pi)^3} \int\frac{d\omega}{2\pi}\frac{e^{i\vec{p}\cdot(\vec{x}_1-\vec{x}_2)}~ e^{-i\omega(t_1-t_2)}}
{\omega^2-\Omega_j^2+i\epsilon},~~j=1,2\label{propamass}
\ee
are the propagators for the mass eigenstates, with
\be
\Omega^2_j =p^2+m^2_j \,. \label{omegaj}
\ee
Here, $E^W_D$ $(E^l_D)$ and $\vec{k}_D$ $(\vec{p}_D)$ label the energy and momentum of the vector boson (charged lepton) at the detection region, while $E^W_S$ $(E^l_S)$ and $\vec{k}_S$ $(\vec{p}_S)$ label those of the production region.\\

The spatial integrals over $\vec{x}_1$ and $\vec{x}_2$ yield the usual spatial momentum conservation $(2\pi)^3\delta(\vp-\vk_D-\vp_D)$ and $(2\pi)^3 \delta(\vk_S-\vp_S-\vp)$,
respectively, leading to an overall momentum conservation in the amplitudes $\mathcal{A}_{\alpha \rightarrow \beta}$ for plane waves. In the transition amplitude for wave packets total momentum conservation is smeared by the wave packet extension. \\

The integrals over the finite time intervals
\be
I = \int_{t_i}^{t_f} dt_1
e^{i(E_D-\omega)t_1}~\int_{t_i}^{t_f} dt_2 e^{-i(E_S-\omega)t_2}
~~;~~E_S= E^W_{\vk_S}-E^l_{\vp_S} ~;~E_D=E_{\vk_D}^W+E^l_{\vp_D} \label{timeints}
\ee
require a careful treatment. In S-matrix theory, $t_i \rightarrow -\infty\,;$ $\, t_f\rightarrow \infty$ and the integrals require an adiabatic switching-on convergence factor. We want to analyze this long-time limit to establish contact with the S-matrix results. It
proves convenient to write
\be
t_f = \frac{T}{2}+ \frac{t}{2},~~ t_i = \frac{T}{2}- \frac{t}{2} \label{Tt}
\ee
and to introduce an adiabatic convergence factor in the integrals
\be
I = \lim_{\delta \rightarrow 0^+}~ e^{i \Delta E\,\frac{T}{2}} ~ \int_{-\frac{t}{2}
}^{\frac{t}{2} }   e^{i(E_D-\omega)t_1}~e^{-\delta |t_1|}~dt_1
~\int_{-\frac{t}{2} }^{\frac{t}{2} }
e^{-i(E_S-\omega)t_2}~e^{-\delta |t_2|}~ dt_2, \label{Iregdel}
\ee
where
\be
\Delta E= E_D-E_S = E_f-E_i, \label{DeltaE}
\ee
and the initial and final energies, $E_i$ and $E_f$, are defined in Eq.~(\ref{inifinener}). We find
\be
\int_{-\frac{t}{2} }^{\frac{t}{2} } e^{i(E -\omega)t_1}~e^{-\delta |t_1|}~dt_1 = 2\pi \, \delta(E-\omega) +2 \, \mathcal{S}\Big[ (E-\omega);t\Big], \label{Sfun}
\ee
where the function (distribution)
\be
\mathcal{S}\Big[ (E-\omega);t\Big] = -\lim_{\delta \rightarrow 0^+}~\int_{\frac{t}{2}}^\infty
\cos\big[(E-\omega)t_1\big] e^{-\delta |t_1|}~dt_1 \label{Sdef}
\ee
has the following behavior
\be
\lim_{t\rightarrow \infty}\mathcal{S}\Big[ (E-\omega);t\Big] = 0 ~~;~~
\lim_{t\rightarrow 0  }\,\mathcal{S}\Big[ (E-\omega);t\Big] = -\pi\,\delta(E-\omega)
 \,.
\label{Sprops}
\ee
For a finite time interval and taking first the limit $\delta \rightarrow 0^+$, it follows from (\ref{Sfun}) that
\be
\pi\, \delta(E-\omega)+ \mathcal{S}\Big[
(E-\omega);t\Big]_{\delta \rightarrow 0^+} \equiv \frac{\sin\Big[
(E-\omega)\frac{t}{2}\Big]}{(E-\omega)} ~~\,. \label{seno}
\ee
We also gather the following useful results
\bea
\lim_{\delta \rightarrow 0^+}~\frac{e^{i( \omega-E+i\delta)\frac{t}{2}}}{i( \omega-E+i\delta)}  & = & - \lim_{\delta \rightarrow 0^+}~\int_{\frac{t}{2}}^\infty e^{i(
\omega-E+i\delta)t_1} dt_1,~~ 
\nonumber \\
\lim_{\delta \rightarrow 0^+}~ \frac{e^{-i(
\omega-E-i\delta)\frac{t}{2}}}{i( \omega-E-i\delta)}  & = & -
\lim_{\delta \rightarrow 0^+}~\int_{\frac{t}{2}}^\infty e^{-i(
\omega-E-i\delta)t_1} dt_1, ~~ 
\label{identities}\eea
which both go to zero at the infinite time limit $t \rightarrow \infty$. \\

We note that the prefactor $e^{i \Delta E\,\frac{T}{2}}$ in (\ref{Iregdel}) combines with the factor $e^{-i(E_f t_f -E_i t_i )}$ in (\ref{amploc}) to yield $e^{-i \overline{E}(t_f-t_i)}$ where $\overline{E} = (E_f+E_i)/2$,  manifestly displaying time translational invariance. \\

The plane wave transition amplitudes are then given by
\be
\mathcal{A}_{\alpha\rightarrow \beta}(t) = -g^2\, \Pi~ (2\pi)^3
 ~\delta(\vk_S-\vp_S-\vk_D-\vp_D) ~U_{\beta j}
~  \mathcal{I}\big[\Omega_j;t\big] ~ U_{j\alpha}\,,
\label{Afinal}
\ee
where $\mathcal{I}\big[\Omega_j;t\big]$ is given by the dispersive integral
\be
\mathcal{I}\big[\Omega_j;t\big] = \frac{2 }{\pi}\int {d\omega} \, \frac{\Big\{\pi\delta(\omega-E_S) +\mathcal{S}\Big[(\omega-E_S);t \Big]
\Big\}\Big\{\pi\delta(\omega-E_D) +\mathcal{S}\Big[(\omega-E_D);t\Big] \Big\}}
{\omega^2-\Omega_j^2+i\epsilon},\label{disI}
\ee
and the momentum argument of the frequencies $\Omega_j$ is $\vp= \vk_D+\vp_D
= \vk_S-\vp_S$. \\

Before we carry out the $\omega$ integral, we note that taking the $t \rightarrow \infty$ limit at this stage yields energy-momentum conservation at each vertex. This yields
\be
\mathcal{I}\big[\Omega_j;t\rightarrow \infty\big] = \frac{2\pi
\, \delta(E_f-E_i)}{E^2_S-\Omega_j^2+i\epsilon}\,,
 \label{Smatx}
\ee
leading to the usual S-matrix result for the transition amplitude for plane wave initial and final states
\be
\mathcal{A}^{(\rm S-mtx)}_{\alpha\rightarrow \beta} = -g^2\, \Pi~ (2\pi)^4
\delta(E_f-E_i)~\delta(\vk_S-\vp_S-\vk_D-\vp_D) ~U_{\beta j} \frac{1}{E^2_S-\Omega_j^2+i\epsilon} U_{j\alpha}\,, \label{ASmtx}
\ee
clearly demonstrating that no oscillations occur for plane waves in the infinite time limit. \\

For finite time interval, the integration over $\omega$ (\ref{disI}) is tedious but straightforward. The result is
\be
\mathcal{I}\big[\Omega_j;t\big] = \frac{D(E_f-E_i;t)}{E^2_S-\Omega_j^2+i\epsilon} +  \mathcal{H}\big[\Omega_j;t \big]   + e^{-i\Omega_j t}  \, \mathcal{F} \big[\Omega_j;t \big], \label{Ifina}
\ee
where to simplify notation we have introduced
\be
D(E_f-E_i;t) = 2\pi \, \delta(E_f-E_i)+2\, \mathcal{S}\Big[(E_f-E_i);t\Big] \,, \label{DofE}
\ee
\bea
\mathcal{H}\big[\Omega_j;t \big] & = & \frac{- i}{2\Omega_j}
\Bigg\{ \frac{e^{i(E_f-E_i +2i\delta)\frac{t}{2}}}{(E_f-E_i
+2i\delta)} \Bigg[\frac{E_f-E_i}{(\Omega_j+E_S)(\Omega_j+E_D)}
\Bigg] \nn \\
& & +  \frac{e^{-i(E_f-E_i -2i\delta)\frac{t}{2}}}{(E_f-E_i
-2i\delta)}\Bigg[\frac{E_f-E_i}{(\Omega_j-E_S)(\Omega_j-E_D)}
\Bigg]\Bigg\} \label{H}
\eea
and
\bea
\mathcal{F} \big[\Omega_j;t \big] &=& \frac{ i}{2\Omega_j}\Bigg\{  \frac{e^{-i(E_D+E_S
-2i\delta)\frac{t}{2}}}{(E_D+E_S -2i\delta)}\Bigg[\frac{ E_D+E_S
}{(\Omega_j+E_S)(\Omega_j+E_D)}\Bigg] \nn \\
& & + \frac{e^{ i(E_D+E_S +2i\delta)\frac{t}{2}}}{(E_D+E_S +2i\delta)}\Bigg[\frac{ E_D+E_S }  {(\Omega_j-E_S)(\Omega_j-E_D)}\Bigg] \Bigg\}.\label{F }
\eea

In the $t\rightarrow \infty$ limit, the first term in (\ref{Ifina}) gives the S-matrix result.
The functions $\mathcal{H}$ and $\mathcal{F} $ only depend on the frequencies $\Omega_j$ in the denominators and vanish in the $ t\rightarrow \infty$ limit, as can be seen from Eq.~(\ref{identities}). It is straightforward to show that there are no poles at $\Omega_j = \pm E_{D,S}$ because the residues vanish, so that both $\mathcal{H}$ and $\mathcal{F}$ must be understood in terms of their principal part.

The disappearance  transition amplitude for \emph{plane waves} are then given by
\bea
\mathcal{A}_{e \rightarrow e} &  = & -g^2 ~\Pi~ (2\pi)^3 \delta(\vk_S-\vp_S-\vk_D-\vp_D) ~\Bigg[\cos^2 \theta ~ \mathcal{I}\big[\Omega_1;t\big] + \sin^2 \theta ~ \mathcal{I}\big[\Omega_2;t\big] \Bigg] \label{Aee}, \nn \\
& &
\\
\mathcal{A}_{\mu \rightarrow \mu} &  = &    -g^2 \Pi~ (2\pi)^3 \delta(\vk_S-\vp_S-\vk_D-\vp_D) ~
\Bigg[\sin^2\theta ~\mathcal{I}\big[\Omega_1;t\big] + \cos^2\theta ~\mathcal{I}\big[\Omega_2;t\big]\Bigg], \label{Amumu} \nn \\
\eea
while the appearance amplitude is given by
\be  \mathcal{A}_{e \rightarrow \mu}= g^2 ~\Pi~ (2\pi)^3 \delta(\vk_S-\vp_S-\vk_D-\vp_D) ~\cos\theta\, \sin \theta ~\Bigg[\mathcal{I}\big[\Omega_1;t\big] - \mathcal{I}\big[\Omega_2;t\big] \Bigg] .\label{Aemu}
\ee

Taking the limit $t\rightarrow \infty$ in these amplitudes, one recovers the S-matrix result obtained by replacing $\mathcal{I}\big[\Omega_j;\infty\big] $ by (\ref{Smatx}), in which case there is no time dependence and no oscillations from interference terms. Instead of taking this limit, we consider a time interval large  compared to microscopic times but of the order of the oscillation time corresponding to the experimental situation in which the baseline is just long enough for a few oscillations. For finite time, we can set $\delta \rightarrow 0^+$ in the above expressions, since we have explicitly separated the delta functions and identified which terms vanish in the formal limit $t\rightarrow \infty$. \\

We would like to emphasize that there is an important distinction between what we do here and the usual S-matrix approach. In the S-matrix approach, the transition amplitude is obtained in the $t\rightarrow \infty$ limit and after taking this limit, one  obtains the transition  probability. In contrast, we obtain the transition amplitude at finite time $t$ and obtain the
probability. \\

In order to extract the most relevant contributions at long time, we consider the ultrarelativistic limit and write
\be
\Omega_1 = \overline{\Omega} - \Delta,~~\Omega_2 = \overline{\Omega} + \Delta,
\label{URO}
\ee
where
\be
\overline{\Omega} = \left[ p^2 +\frac{m^2_1+m^2_2}{2}\right]^\frac{1}{2},~~ \Delta= \frac{\delta m^2}{4\overline{\Omega}},~~\delta m^2 = m^2_2-m^2_1,  \label{diffs}
\ee
taking $\Delta \ll \oO$ as is the case for ultrarelativistic neutrinos. We can then write
\bea
\mathcal{H}\big[\Omega_j;t \big] &  =  &  \mathcal{H}\big[\overline{\Omega};t \big] + (-1)^j
\Bigg[\overline{\Omega}~\frac{d \mathcal{H}\big[\overline{\Omega};t\big]}{d\overline{\Omega}}
\Bigg]~\Bigg(\frac{\delta m^2}{4\overline{\Omega}^2}\Bigg), \nonumber \\\mathcal{F} \big[\Omega_j;t \big] &  =  & \mathcal{F} \big[\overline{\Omega};t \big] + (-1)^j \Bigg[\overline{\Omega}~\frac{d \mathcal{F} \big[\overline{\Omega};\frac{t}{2}\big]}{d\overline{\Omega}} \Bigg]~
\Bigg(\frac{\delta m^2}{4\overline{\Omega}^2}\Bigg)\label{expansions}.
\eea
We consider the realistic situation in which $\delta m^2/\overline{\Omega}^{\,2} \ll 1$ and keep the terms of $\mathcal{O}(\delta m^2/\overline{\Omega}^{\,2})$ \emph{only} in the S-matrix term and in the exponentials $e^{i\Omega_{1,2}t}$, neglecting the small corrections both in $\mathcal{H}$ and $\mathcal{F}$. The reason for keeping the $\mathcal{O}(\delta m^2/\overline{\Omega}{\,^2})$ correction in the S-matrix contribution will become clear below when we discuss the appearance probability. \\

We can further define
\be
F[\oO;t] = e^{-i {\oO} t} \mathcal{F} \big[\oO;t] ,
\ee
thus simplifying the expressions for the disappearance and appearance amplitudes
\bea
\mathcal{A}_{e \rightarrow e} &=& -g^2 ~\Pi~ (2\pi)^3\, \delta(\vk_S-\vp_S-\vk_D-\vp_D) ~  \Bigg\{ D(E_f-E_i;t) \Bigg[ \frac{\cos^2\theta}{E^2_S-\Omega^2_1+i\epsilon}+ \frac{\sin^2\theta}{E^2_S-\Omega^2_2+i\epsilon} \Bigg]  \nonumber \\ & &+\,
\mathcal{H}\big[\oO;t \big]  +F[\oO;t]\Big[\cos^2\theta ~ e^{ i \frac{\delta m^2}{4\oO} t}
+\sin^2 \theta ~ e^{ -i \frac{\delta m^2}{4\oO} t}  \Big] \Bigg\} + \mathcal{O}\Bigg( \frac{\delta m^2}{\oO^2} \Bigg), \label{Aee2}
\eea
\bea
\mathcal{A}_{e \rightarrow \mu}    &=&   g^2 ~\Pi~ (2\pi)^3 \,\delta(\vk_S-\vp_S-\vk_D-\vp_D) ~   \cos \theta \sin \theta  \Bigg\{  D(E_f-E_i;t) \Bigg[ \frac{1}{E^2_S-\Omega^2_1+i\epsilon}- \frac{1}{E^2_S-\Omega^2_2+i\epsilon} \Bigg] \nonumber \\
& & + \, 2i F[\oO;t]  ~    \sin \Big[\frac{\delta m^2}{4\oO}\,t \Big] \Bigg\}+ \mathcal{O}\Bigg( \frac{\delta m^2}{\oO^2} \Bigg)\,.
\label{Aemu2}
\eea
The amplitude $\mathcal{A}_{\mu \rightarrow \mu}$ can be obtained from (\ref{Aee2}) by the replacement $ \sin^2\theta  \leftrightarrow \cos^2 \theta$. \\

The expressions above clearly exhibit how and where oscillatory interference terms arise in the probabilities. The usual S-matrix result is obtained in the $t\rightarrow \infty $ limit where $D(E_f-E_i;t)\rightarrow 2\pi\,\delta(E_f-E_i)$ and $F[\oO,t], \mathcal{H}[\oO,t] \rightarrow 0$.  It is clear that in this limit, the oscillatory behavior is suppressed and no
interference terms can possibly survive in the transition probabilities. \\


In obtaining the transition probabilities, we recognize two types of oscillatory terms: terms that feature exponentials of the form $e^{ \pm i(E_D \pm E_S)t}$ and those of the form $e^{  \pm i\delta m^2 t/4\oO}$. The former are fast oscillatory terms on \emph{microscopic} time scales, whereas the latter are slow phases on these time scales and only manifest themselves on much longer time scales, of order of the baseline $t \sim L$. The fast oscillatory terms average out on the (much) longer time scale. \emph{After} obtaining the transition probabilities at a \emph{finite time} $t$, we discard terms that feature the fast oscillations   that average out in  the long time limit, akin to what happens in the ``rotating wave approximation'' in quantum optics \cite{QO}, and finally take the $\delta \rightarrow 0$ limit, obtaining the \emph{plane wave} transition probabilities
\bea
\mathcal{P}_{e \rightarrow e}  & = & g^4~\Pi^2 ~V~ \delta(\vk_S-\vp_S-\vk_D-\vp_D)\Bigg\{ 2\pi\,t~\delta(E_f-E_i)~\Bigg|\frac{\cos^2\theta}{E^2_S-\Omega^2_1+i\epsilon}+ \frac{\sin^2\theta}{E^2_S-\Omega^2_2+i\epsilon}  \Bigg|^{\,2}\nonumber \\
&& + \,\frac{1}{4\oO^2}\Bigg[ \frac{1}{(\oO+E_S)^2(\oO+E_D)^2} + \frac{1}{(\oO-E_S)^2(\oO-E_D)^2} \Bigg]\Bigg(1- \sin^2 2\theta ~\sin^2\Big[ \frac{\delta m^2}{4\oO}\,t\Big] \Bigg)\nonumber \\ & & + \,\frac{1}{4\oO^2}\Bigg[ \frac{1}{(\oO+E_S)^2(\oO+E_D)^2} + \frac{1}{(\oO-E_S)^2(\oO-E_D)^2} \Bigg]\Bigg\},   \label{Pee} \\
\mathcal{P}_{e \rightarrow \mu}  &=& g^4 ~\Pi^2~V~ (2\pi)^3 \delta(\vk_S-\vp_S-\vk_D-\vp_D) ~ \sin^2 2\theta \nonumber \\
&&\times \, \Bigg\{ \frac{(2\pi)}{4}~t~\delta(E_f-E_i)~ \Bigg| \frac{1}{E^2_S-\Omega^2_1+i\epsilon}- \frac{1}{E^2_S-\Omega^2_2+i\epsilon} \Bigg|^{\,2} \nonumber \\   && + \,
\frac{1}{4\oO^2}\Bigg[ \frac{1}{(\oO+E_S)^2(\oO+E_D)^2} + \frac{1}{(\oO-E_S)^2(\oO-E_D)^2} \Bigg]  ~\sin^2\left[ \frac{\delta m^2}{4\oO}\,t\right] \Bigg\} . \label{Pemu}
\eea
Here, we have neglected contributions of $\mathcal{O}(\delta m^2/\oO^2)$. As before, the disappearance probability $\mathcal{P}_{\mu\rightarrow \mu}$ is obtained from (\ref{Pee}) by
the substitution $\cos^2\theta \leftrightarrow \sin^2 \theta$. \\

The first line in the above expressions is the S-matrix result, where we have used $(2\pi \delta (E_f-E_i))^2 \rightarrow (2\pi)~t~\delta(E_f-E_i)$ as usual. The second line in (\ref{Pee}) arises from $|\mathcal{H}(\oO;t)|^2$. This term is a direct result of calculating the probability at \emph{finite time} before taking the $t\rightarrow \infty $ limit. The two procedures \emph{do not commute}. On one hand, taking the $t\rightarrow \infty$ limit first results in the vanishing of $\mathcal{H}$ by the averaging of the oscillatory terms (Riemann-Lebesgue lemma). On the other hand, obtaining the probability first and taking the long time limit afterward yields the contribution from the modulus squared of each oscillatory term, leading to the second line in (\ref{Pee}). \\

Writing $\Omega_{1,2}$ as in Eqs. (\ref{URO},\ref{diffs}), it follows that the S-matrix contribution to the appearance probability is
\be
\mathcal{P}^{\rm S-mtx}_{e \rightarrow \mu} \propto  \delta(E_f-E_i)\Big[\frac{\delta m^2\,t}{4\oO} \Big] \frac{\delta m^2\oO}{(E^2_S-\oO^2)^4}. \,
\label{PemuSmatrix}
\ee
Upon integrating on the final density of states for times such that there are few oscillations, namely $\delta m^2 t/4\oO \sim 1$, the S-matrix contribution is \emph{much smaller} than the oscillatory terms in the second line of (\ref{Pemu})  because $\delta m^2 \ll \oO^{\,2}, E^2_{S,D}$. On the other hand, the S-matrix contribution to the disappearance probability is
\be
\mathcal{P}^{S-mtx}_{e \rightarrow e} \propto  t\,\delta(E_f-E_i)  \frac{1}{(E^2_S-\oO^2)^2} \, ,
\label{PeeSmatrix}
\ee
which upon integration over the density of states will dominate over the oscillatory terms in these probabilities for $t \sim t_{osc} \sim 4\oO/\delta m^2$. Thus we see that appearance and disappearance probabilities are \emph{fundamentally} different. For time scales during which oscillation phenomena can be observed, the S-matrix contribution to the disappearance      probability dominates while this   may not  be the case for the appearance probability, as the oscillatory terms emerging at finite time may be comparable to or larger than the S-matrix contributions.\\

Lastly, we would also like to note that the oscillatory terms feature an energy dependence very different from that of the S-matrix contribution.

\subsection{Which time scale?}\label{sec:look}
An important question emerges from the analysis above. \emph{What is the value of the time $t$ in these probabilities given in Eqs.~(\ref{Pee}) and (\ref{Pemu})}? The linear time dependence, a consequence of \emph{total energy} conservation, also emerges in many other analyses with wave packets \cite{rich,grimus1,grimus2,grimus3,grimus4,grimus5,giunti1,giunti2,giunti3,giunti4,giunti5,giunticohe,pila,cardall,dolgov1,dolgov2,dolgov3,keister,akhme1,akhme2}, from which the S-matrix evaluation of the transition probabilities always yields a result proportional to $T/L^2$ where $T$ is the ``total reaction time'' arising from squaring the delta function associated with overall energy conservation and $L$ is the baseline. The factor $1/L^2$ has a clear physical meaning corresponding to the ratio of the neutrino flux received at the far detector and that produced at the near detector. In the plane wave limit, the factor $1/L^2$ disappears because we lose the localization property of wave packets; however, the time $t$ in the probabilities (\ref{Pee}) and (\ref{Pemu}) plays the same role as the linear time dependence $T$ in standard S-matrix calculations. This is clear shown in Eqs.~(\ref{PemuSmatrix}) and (\ref{PeeSmatrix}) denoting the S-matrix contributions to the probabilities given in (\ref{Pee}) and (\ref{Pemu}). However, the finite time corrections to the appearance and disappearance probabilities in (\ref{Pee}) and (\ref{Pemu}) do not feature the linear time dependence. This observation exactly motivates us to address the question what the time scale $t$ really means. \\

In a typical S-matrix calculation   one divides by $T$, the ``total reaction time,''
to obtain a \emph{transition rate}. However, this interpretation needs revision in the case of long-baseline experiments. In these experiments, a  charged lepton is measured at the near detector (source) whereas another charged lepton is measured at the far detector. The intermediate neutrino state propagates between these as a wave packet.  Therefore, there are two time scales in this case:  the time at which the near-detector measurement of the charged lepton occurs and the time at which the detection of the charged lepton at the  far detector occurs.  The question of what precisely is the time $t$ in the S-matrix calculation is in principle independent from the wave packet treatment and is an inherent question to the S-matrix formulation of the production, detection and propagation in long-baseline experiments. \\

There is another problem with the analysis that we have done so far. In the absence of oscillations, dividing by time or taking the time derivative yields identical results. However, in the presence of oscillatory contributions, the transition \emph{rate} must be obtained by taking the \emph{time derivative}. The time derivative of the oscillatory terms featured in the transition probabilities (\ref{Pee}-\ref{Pemu}) are of order $\delta m^2/\oO$. For the disappearance probabilities, this derivative term is subleading with respect to the S-matrix contribution, but it may be of the same order or larger in the case of the appearance probability, as discussed above. \\

   The question of how to interpret the total time $t$ in the transition probabilities, along with the conceptual
      differences between appearance and disappearance (insofar as the oscillatory contributions), suggests a re-examination of how these
 probabilities should be calculated. \\

\begin{figure}[h!]
\begin{center}
\includegraphics[width=12cm,keepaspectratio=true]{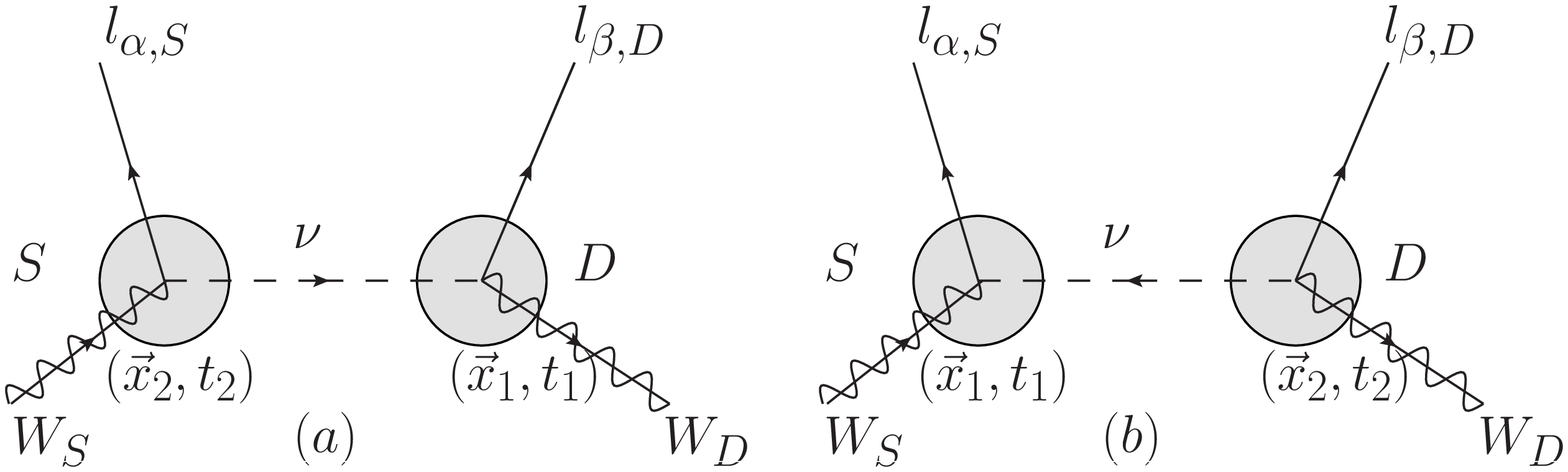}
\caption{The two contributions to the transition amplitude
(\ref{2ndord}) from the two different Wick contractions. For both
diagrams $t_1 > t_2$. } \label{fig:vertices}
\end{center}
\end{figure}

It proves illuminating to understand the result above from
``old-fashioned'' time dependent perturbation theory with a finite
time interval $t_f,t_i$. To this end, we will need the second order matrix element
\bea
\mathcal{A}_{\alpha \rightarrow \beta} &=&
 -g^2 \int d^3 x_1   d^3 x_2
 \int_{t_i}^{t_f}dt_1 \int_{t_i}^{t_1}dt_2 \nonumber \\ &&
 \langle {W} (\vk_D), {l}_{\beta}(\vp_D); {l}_{\alpha}(\vp_S) |W(x_1,t_1)\, l(x_1,t_1)\,\nu(x_1,t_1)
\, W(x_2,t_2)\,l(x_2,t_2)\nu(x_2,t_2)| {W} (\vk_S)\rangle, \nonumber \\
 \label{2ndord}
\eea
where we have suppressed the flavor indices in the interaction for simplicity of notation. We note that in the above
 expression time is ordered $t_f \geq t_1 > t_2$. There are two Wick
 contractions corresponding to the processes displayed in Fig.
 (\ref{fig:vertices}). In diagram (\ref{fig:vertices}a) the
 charged lepton  at the near detector $l_S$ is created at $t_2 < t_1$, namely \emph{before} the
 charged lepton at the far detector $l_D$, whereas Fig.
 (\ref{fig:vertices}b) displays the opposite process in which the
 charged lepton at the far detector is created \emph{before} the charged lepton at the near detector.
\bea
\mathcal{A}^{(a)}_{\alpha \rightarrow \beta}  & = & -g^2 ~\Pi~
(2\pi)^3\, \delta(\vk_S-\vp_S-\vk_D-\vp_D) ~e^{i \Delta
E\,\frac{T}{2}} \times \nonumber \\ & &
\frac{U_{\alpha,j}}{2\Omega_j(q)}\Bigg\{\frac{~ {2i}
~\sin\big(\Delta E \frac{t}{2}\big)}{\big(E_s-\Omega_j(q)\big)\Delta
E} + \frac{e^{-i\Delta E \frac{t}{2}}-e^{-i\Omega_j t}e^{ i(E_D+E_S
 )\frac{t}{2}} }{(\Omega_j(q)-E_S)(\Omega_j(q)-E_D)} \Bigg\} \,U_{j,\beta}\,, \label{Aaba}
\eea
\bea
\mathcal{A}^{(b)}_{\alpha \rightarrow \beta}  & = & -g^2 ~\Pi~
(2\pi)^3 \,\delta(\vk_S-\vp_S-\vk_D-\vp_D) ~e^{i \Delta
E\,\frac{T}{2}} \times \nonumber \\ & &
\frac{U_{\alpha,j}}{2\Omega_j(q)}\Bigg\{\frac{~ -{2i}
~\sin\big(\Delta E \frac{t}{2}\big)}{\big(E_s+\Omega_j(q)\big)\Delta
E} + \frac{e^{i\Delta E \frac{t}{2}}-e^{-i\Omega_j t}e^{ -i(E_D+E_S
 )\frac{t}{2}} }{(\Omega_j(q)+E_S)(\Omega_j(q)+E_D)} \Bigg\} \,U_{j,\beta}\,, \label{Aabb}
\eea
where $\vq = \vk_S-\vp_S$. It is straightforward to confirm  that $\mathcal{A}^{(a)}_{\alpha
\rightarrow \beta}+\mathcal{A}^{(b)}_{\alpha \rightarrow \beta}$ coincides with the results (\ref{Afinal},\ref{Ifina}) at finite time taking $\delta,\epsilon = 0$.\\

In process $(a)$, the interaction vertex annihilates the $W_S$
in the initial state and creates both a neutrino and the charged
lepton $l_{\beta,S}$ at the space-time coordinates $(\vec{x}_2,t_2)$
in a superposition of product Fock states, namely in an
\emph{entangled state} \cite{glashow}. In process $(b)$, the initial
$W_S$ and the final lepton $l_{\beta,S}$ are annihilated and created
at $(\vec{x}_1,t_1)$ respectively and the neutrino and
$W_D,l_{\alpha,D}$ are created at $(\vec{x}_2,t_2)$, with the time
ordering $t_1 > t_2$. The combination of both contributions yields
the time ordered product as usual. However, for a long-baseline
experiment when the size of the near and far detectors are much
smaller than the distance between them, these processes should
\emph{not} contribute on equal footing: the charged lepton produced
at the source (near detector) will be detected \emph{much earlier}
than the charged lepton produced at the far detector. \\

A  somewhat extreme example which can illustrate the point is SN1987a in the Large Magellanic Cloud.  The charged leptons
produced in the explosion along with the concomitant neutrinos are
trapped, or ``detected,'' in the optically thick medium, whereas the
neutrinos produced at the source are detected at a distance $\sim
51$ kpc away from the production region. The detection of the
charged lepton in the near detector \emph{disentangles} the quantum
state. This suggests that long-baseline experiments
actually involve \emph{two} times: the time at which the charged
lepton produced with the neutrino is detected in the near detector
and the time at which the charged lepton produced by the neutrino is
detected in the far detector. Obviously, only the process described
by the amplitude $\mathcal{A}^{(a)}_{\alpha \rightarrow \beta}$ above describes this physical situation. However, this
amplitude does \emph{not} describe  the process of measurement of
the charged lepton at the near detector, namely, the
\emph{disentanglement} of the charged lepton and the neutrino. The
correct description requires addressing the issue of coherence and
entanglement.

\section{Coherence, entanglement and oscillations}\label{sec:entanglement}

      In order to study aspects of coherence we consider a simplified
      interaction Lagrangian density \be \mathcal{L}_I = g\, W \,e
      \,\nu_e = g \,W \,e (\cos \theta\, \nu_1 +\sin \theta \, \nu_2),
      \label{lagsim}\ee focusing only on one lepton, which we refer
      to as the ``electron'' to simplify the discussion. The full
      coupling as in Eq. (\ref{Interaction}) can be treated similarly
      without modifying the main conclusions. Although $W$ may be interpreted as a charged vector boson,
      the analysis is obviously the same if it describes a pion
      field. \\

We can study aspects of coherence by focusing on the Fock state obtained upon evolution of the decaying initial state. We consider a plane-wave Fock initial state $\big|W(\vk)\big\rangle$ at $t_i=0$. By time $t$, this initial state has evolved into  $\big|W(\vk)\big\rangle ~e^{-iE^W_{\vk}t}+ \big|\Psi_e(t)\big\rangle$. To lowest order in the interaction we find the second term to be
\be
\big|\Psi_e(t)\big\rangle =  i g~e^{-iH_0t}~ \int_0^t dt'\int d^3x ~\Big[W(\vec{x},t')\, e(\vec{x},t') (\cos \theta\, \nu_1(\vec{x},t')+ \sin \theta\,
    \nu_2(\vec{x},t'))\Big]~\big|W(\vk)\big\rangle, \label{newstate}
\ee
where all the fields are in the interaction picture. Though the  field
operator $W$ can either annihilate the initial state or create another $W$ particle,  the state with two $W$ particles features faster oscillations that will average out. In what follows, we consider only the Fock state resulting from the annihilation, leading to the state
\bea
\big|\Psi_e(t)\big\rangle & \simeq &
\frac{g}{2\sqrt{2VE^W_{\vk}}}~e^{-iE^W_{\vk}t}~\sum_{\vq}    \Bigg\{
\frac{\sin\theta}{\sqrt{\Omega_{2, \vp} ~ E^{e}_{\vq}}}~\Big|\nu_{2,\vp}\rangle \Big|e_{\vq} \rangle \Bigg[\frac{e^{ i(E^W_{\vk}-E^{e}_{\vq}-\Omega_{2,\vp} )t}-1}{(E^W_{\vk}-E^e_{\vq}-\Omega_{2,\vp} )} \Bigg]
\nonumber \\ & + & \frac{\cos\theta}{\sqrt{ \Omega_{1,\vp}~
E^e_{\vq}}}~\Big|\nu_{1,\vp}\rangle \Big|e_{\vq} \rangle
\Bigg[\frac{e^{ i(E^W_{\vk}-E^e_{\vq}-\Omega_{1,\vp} )t}-1}{ (E^W_{\vk}-E^e_{\vq}-
 \Omega_{1,\vp} )} \Bigg] \Bigg\}~~;~~ \vp = \vk-\vq \label{finstate}
\eea
in which the electron and the neutrinos are \emph{entangled}\footnote{The result for the wavefunction in Ref. \cite{glashow} may be understood using a (non-perturbative) Wigner-Weisskopf approximation for the decaying parent particle, replacing $E_W \rightarrow E_W-i\Gamma_W$. Taking $t \gg 1/\Gamma_W$ in the integral replaces the brackets in (\ref{finstate}) by $1/(E_W-E^e-\Omega_j -i\Gamma_W)$ whose absolute value is proportional to $\delta(E_W-E^e-\Omega_j )/\Gamma_W$.}. The neutrino state that is entangled with the
muon is obtained from (\ref{finstate}) by replacing $\cos \theta \rightarrow -\sin \theta~;~ \sin\theta \rightarrow  \cos\theta$.

\subsection{Unobserved daughter particles: time evolution of the density matrix}

If the electrons (or daughter particle in Ref.(\cite{glashow})) are not observed, they can be
traced out of the \emph{density matrix} $ \big|\Psi_e(t)\big\rangle \big \langle \Psi_e(t)\big|$. This gives the \emph{reduced} density matrix
\bea
\rho_r(t)& =& \mathrm{Tr}_{e}\big|\Psi_e(t)\big\rangle \big \langle \Psi_e(t)\big| \nonumber \\
 & = &   \frac{g^2}{8 V E^W_{\vk}} \sum_{\vq}    \Bigg\{
\frac{\sin^2\theta}{\Omega_{2, \vp} ~ E^e_{\vq}} \, \Big|\nu_{2,\vp}\rangle \langle\nu_{2,\vp}\Big| ~
\Bigg[\frac{\sin\Big(\big(E^W_{\vk}-E^e_{\vq}-\Omega_{2,\vp}  \big)\frac{t}{2}\Big)}{\big(E^W_{\vk}-E^e_{\vq}-\Omega_{2,\vp}\big)/2 }
\Bigg]^2 \nonumber \\ && + \,  \frac{\cos^2\theta}{\Omega_{1, \vp} ~ E^e_{\vq}}\,\Big|\nu_{1,\vp}\rangle \langle \nu_{1,\vp}\Big| ~\Bigg[\frac{\sin\Big(\big(E^W_{\vk}-E^e_{\vq}-\Omega_{1,\vp}  \big)\frac{t}{2}\Big)}{\big(E^W_{\vk}-E^e_{\vq}-\Omega_{1,\vp}\big)/2 }
 \Bigg]^2 \nonumber \\ & &+ \, \frac{\sin2\theta}{2E^e_{\vq}\, \sqrt{\Omega_{2, \vp}~\Omega_{1,
     \vp}}}\Bigg[\frac{\sin\Big(\big(E^W_{\vk}-E^e_{\vq}-\Omega_{2,\vp}  \big)\frac{t}{2}\Big)}{\big(E^W_{\vk}-E^e_{\vq}-\Omega_{2,\vp}\big)/2 }
     \Bigg]\Bigg[\frac{\sin\Big(\big(E^W_{\vk}-E^e_{\vq}-\Omega_{1,\vp}  \big)\frac{t}{2}\Big)}{\big(E^W_{\vk}-E^e_{\vq}-\Omega_{1,\vp}\big)/2 }
     \Bigg]\nonumber \\ & &\times \, \Bigg[ e^{ - i \frac{\delta m^2}{4\oO} t} \, \Big|\nu_{2,\vp}\rangle \langle
    \nu_{1,\vp}\Big|+ e^{  i \frac{\delta m^2}{4 \oO} t} \, \Big|\nu_{1,\vp}\rangle \langle
    \nu_{2,\vp}\Big|\Bigg] \Bigg\}~~;~~\vp = \vk-\vq\,.
    \label{redrho}
\eea
This expression contains remarkable information. The function
$\sin^2(xt)/x^2$ is the usual ``diffraction'' function of Fermi's
Golden rule. In the formal long time limit $\sin^2(xt)/x^2
\rightarrow \pi\,t\,\delta(x)$, the first two terms of the density
matrix, which are the diagonal entries in the mass basis, describe the production process
of the mass eigenstates. As will be seen below, in the long time
limit, the time derivative of these two terms yields the production
rate  for the mass eigenstates. In the formal $t\rightarrow \infty$ limit, these
are the diagonal terms obtained in Ref. \cite{glashow}, where in that
reference, the product of delta functions is again understood as the
total time elapsed times an energy conserving delta function. \\

\begin{figure}[h!]
\begin{center}
\includegraphics[height=4cm,keepaspectratio=true]{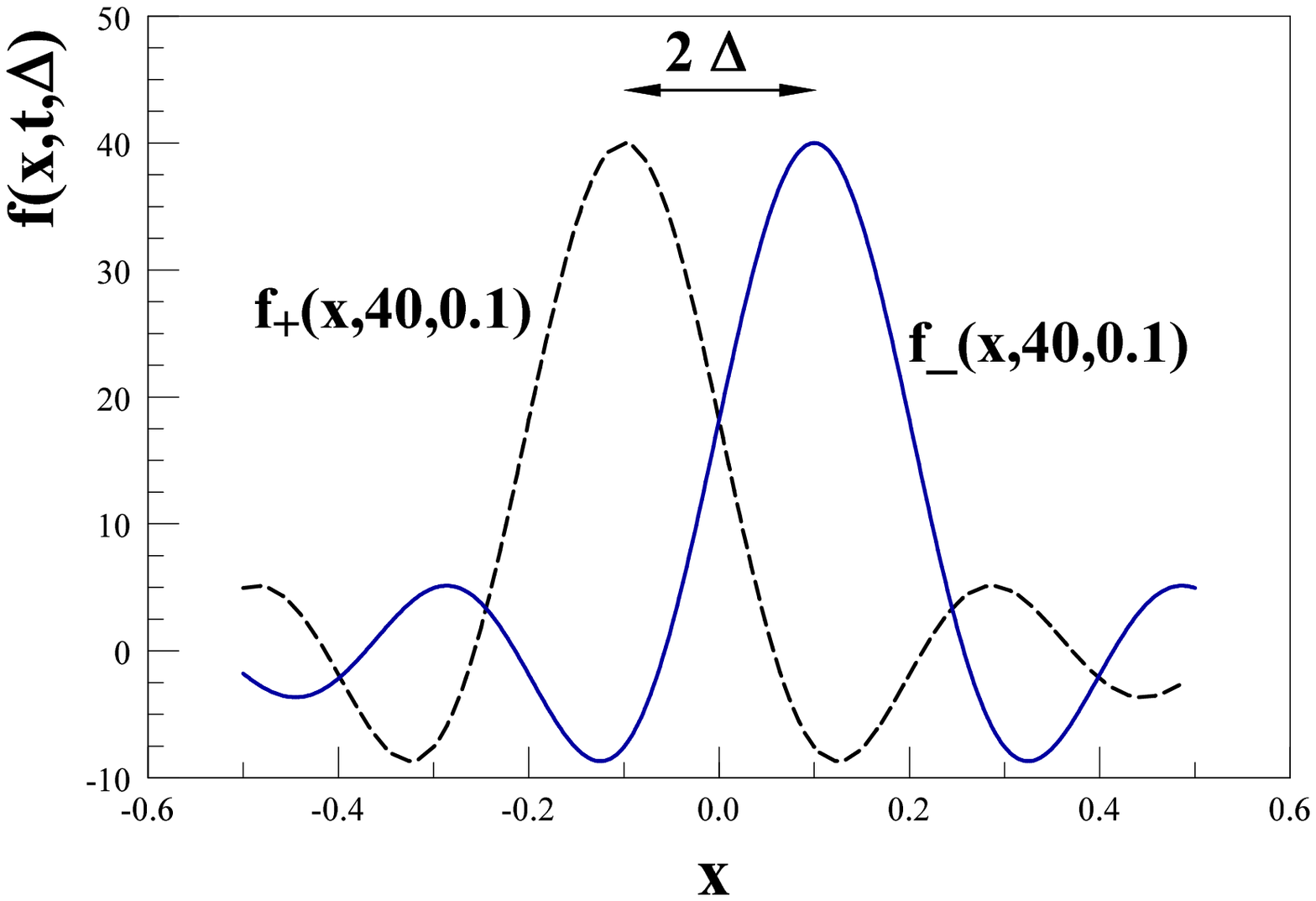}
\includegraphics[height=4cm,keepaspectratio=true]{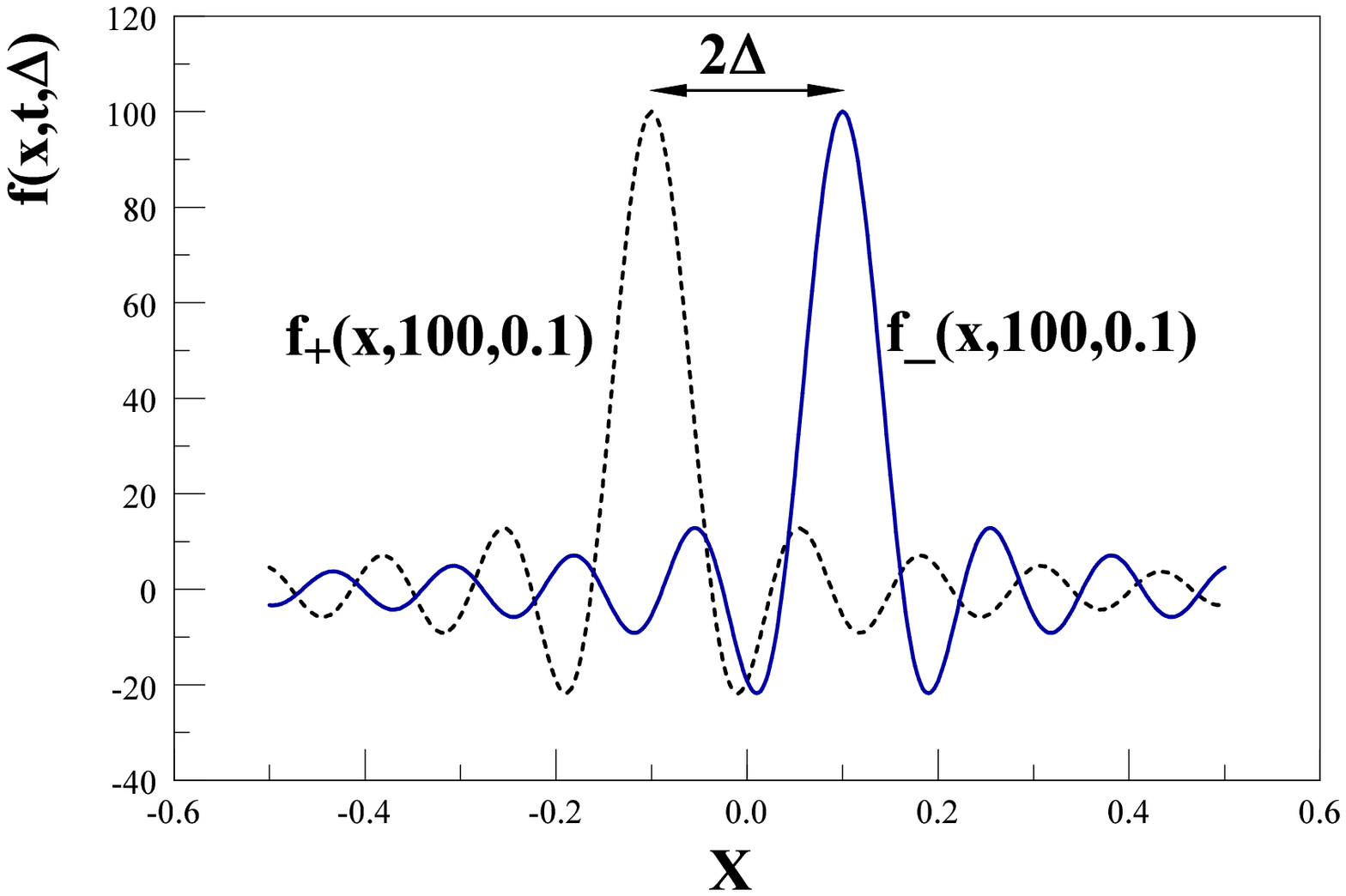}
\caption{The functions $f_\pm(x,t,\Delta)$ vs. $x$ for
$t=40,100$, $\Delta=0.1$} \label{fig:fpm}
\end{center}
\end{figure}

\begin{figure}[h!]
\begin{center}
\includegraphics[height=4cm,keepaspectratio=true]{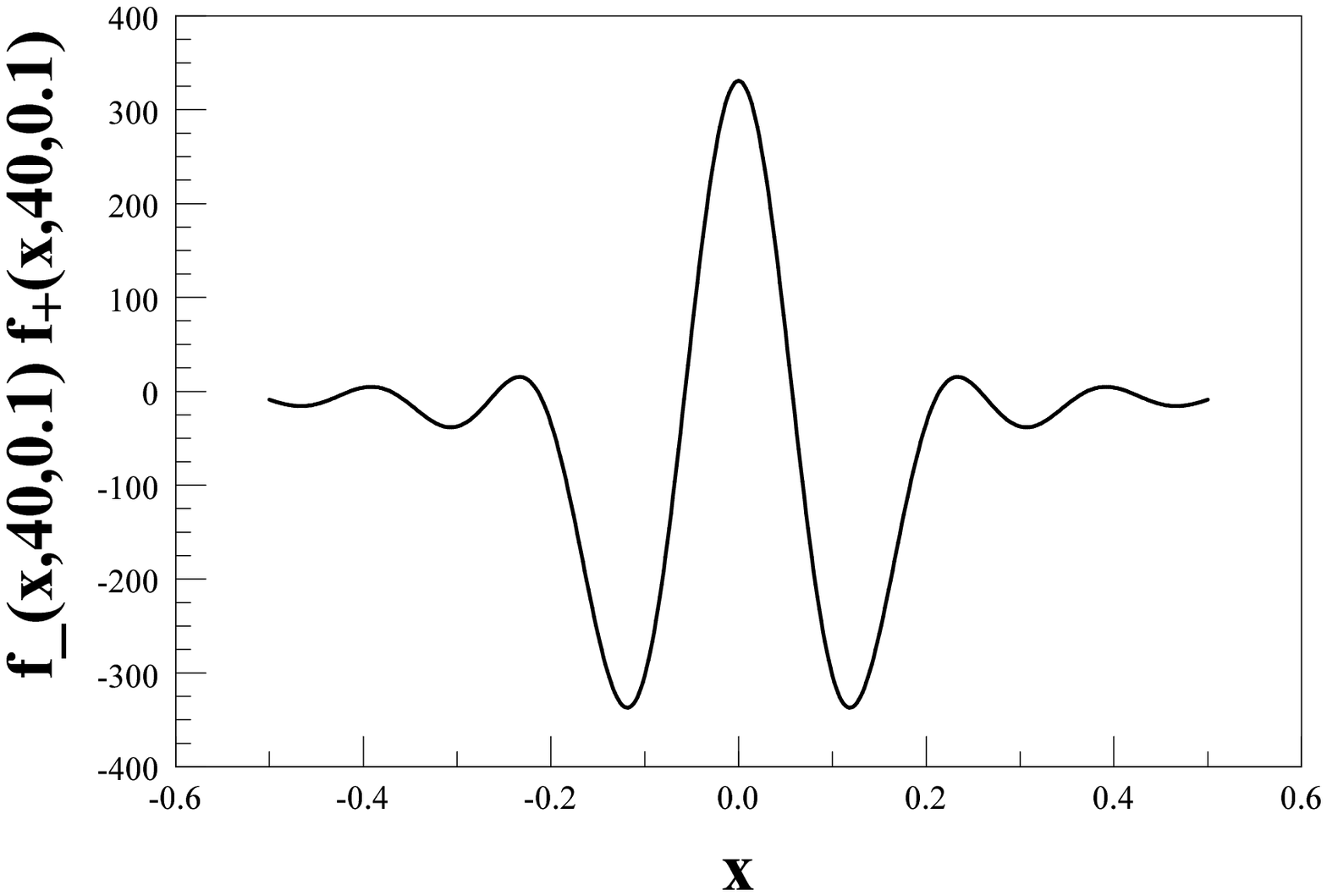}
\includegraphics[height=4cm,keepaspectratio=true]{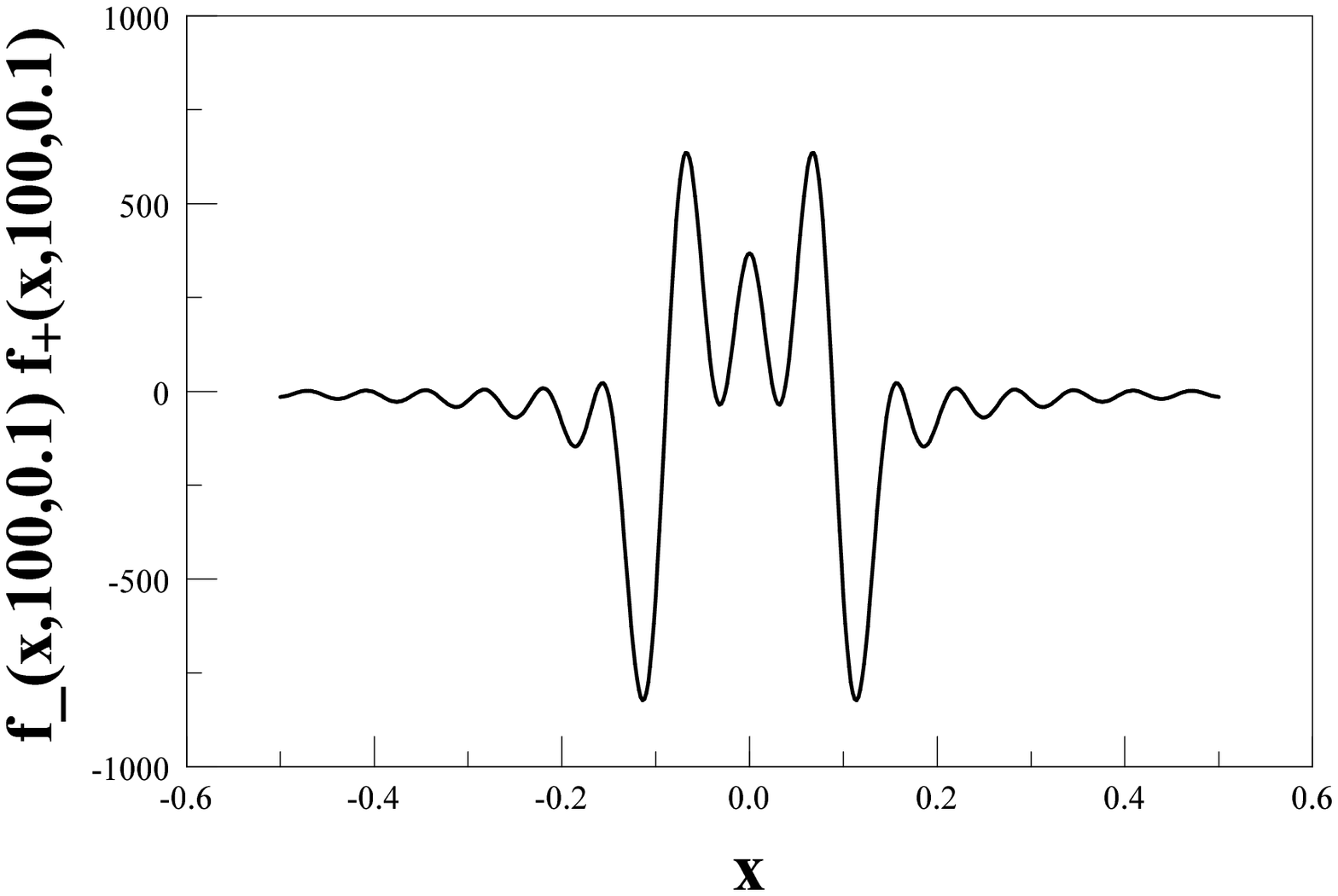}
\caption{The products $f_-(x,t,\Delta)f_+(x,t,\Delta)$ vs. $x$ for
$t=40,100$,$\Delta=0.1$} \label{fig:fprods}
\end{center}
\end{figure}

The off-diagonal terms in the last line of (\ref{redrho}) describe
the ``coherences'' and display the oscillatory phases from the
interference of the mass eigenstates. The time dependent factors
of the off-diagonal density matrix elements determine the
\emph{coherence} between the mass eigenstates and will be a
ubiquitous contribution in the real time description of oscillations
that follows below. The functions \be f_{\pm}(x,t;\Delta) = \frac{2
\sin\Big[\big(x\pm \Delta\big)\frac{t}{2}\Big]}{\big(x\pm
\Delta\big)} ~~;~~ x=E^W_{\vk}-E^e_{\vq}-\oO_{\vp} ~~;~~ \Delta =
\frac{\delta m^2}{4\overline{\Omega}}\,,  \label{fpm}\ee as functions
of $x$ are strongly peaked at $x\pm \Delta=0$ with height $t$ and
width $\sim 2\pi/t$. In the infinite time limit $f_\pm(x,t,\Delta)
\rightarrow 2\pi \,\delta(x\pm\Delta)$ and thus, their product would
vanish in this limit, leading to the vanishing of the coherence. This is the result obtained in Ref.~\cite{glashow}. However, \emph{at finite time }$t$, they feature
a non-vanishing overlap when $2\Delta \lesssim 2\pi/ t$. We
recognize this precisely as the condition for oscillations. We note
that $t \sim \pi/\Delta$ yields a \emph{macroscopically} large
time scale. The functions $f_\pm(x,t,\Delta)$ and their products are
depicted in Figs. (\ref{fig:fpm},\ref{fig:fprods}) for the values
$\Delta=0.1$, $t=40,100$, respectively.

It is straightforward to find
\bea
f_-(x,t,\Delta)f_+(x,t,\Delta) & =  & \frac{\sin(\Delta\,t)}{\Delta}\Bigg[\frac{\sin[(x-\Delta)t]}{(x-\Delta)}+
 \frac{\sin[(x+\Delta)t]}{(x+\Delta)}\Bigg]   \nonumber\\ & & + \, \frac{\cos(\Delta\,t)}{\Delta}\Bigg[\frac{\sin^2[(x-\Delta)\frac{t}{2}]}{(x-\Delta)}- \frac{\sin^2[(x+\Delta)\frac{t}{2}]}{(x+\Delta)}\Bigg]
\label{formula} \,.
\eea
In the long time limit, the terms in the first bracket yield a sum of delta functions at $x = \pm \Delta$, whereas the second terms are similar to a principal part. Upon integrating the product of $f_-f_+$ with functions of compact support, the contribution from the second line in (\ref{formula}) is negligible in the long time limit. Therefore, the long time limit of $f_+f_-$ can be replaced by
\be
f_-(x,t,\Delta)f_+(x,t,\Delta)= \pi \, \frac{\sin(\Delta\,t)}{\Delta}\Bigg[\delta(x-\Delta) + \delta(x+\Delta) \Bigg] \label{ltff}
\ee
For a large time $t\ll 2\pi/\Delta$, it follows that the product
$f_-(x,t,\Delta)f_+(x,t,\Delta)\sim \pi\, t [\delta(x-\Delta) +
\delta(x+\Delta)] $ grows linearly in time but is bound in time.  For $t > 2\pi /\Delta$, it
oscillates with frequency $2\pi/\Delta$. Therefore, we conclude that upon integration with a smooth density of
states, the off-diagonal terms in the density matrix grow linearly in
time for $t \ll t_{osc} = 2\pi/\Delta$, but feature a bound
oscillatory behavior of frequency $\Delta$ for $t \gtrsim
2\pi/\Delta$. \\

\begin{figure}[h!]
\begin{center}
\includegraphics[width=4in,keepaspectratio=true]{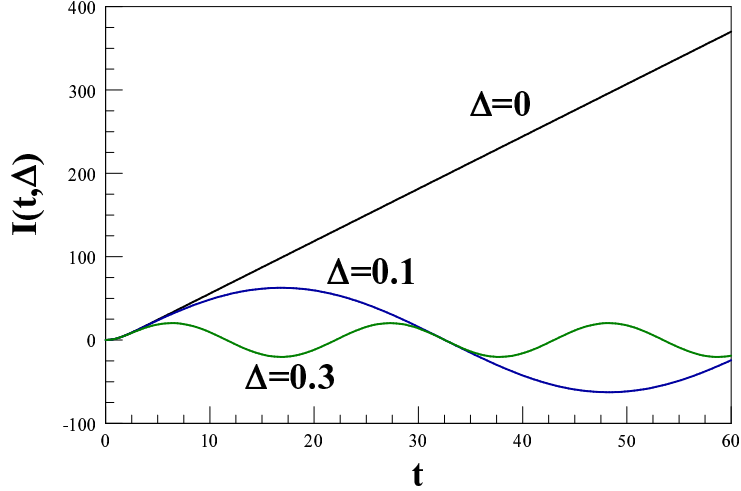}
\caption{The integral $I(t,\Delta) = \int^\infty_{-\infty} e^{-x^2}
f_+(x,t,\Delta)f_-(x,t,\Delta) dx $ vs. $t$ for
$\Delta=0,0.1,0.3$ } \label{fig:integral}
\end{center}
\end{figure}

Whereas the diagonal terms,  {\it i.e.}, the  first two terms in the reduced density
matrix (\ref{redrho}), are proportional to $4\sin^2(xt)/x^2
\rightarrow 2\pi ~t \delta(x)$, the coherences or the off diagonal terms
are linear in time and of the \emph{same order} as the diagonal
elements for $t \leq  t_{os} = 2\pi/\Delta$, but are of
$\mathcal{O}(1/\Delta t)$ and oscillate fast compared to the
diagonal terms for $t \gg t_{osc}$. This is similar to the
phenomenon observed in the transition probability in the previous
section.   This behavior is displayed in Fig. (\ref{fig:integral}),
where as an example we consider a smooth density of states and   the
integral
\be
I(t,\Delta) = \int^\infty_{-\infty} e^{-x^2}
f_+(x,t,\Delta)f_-(x,t,\Delta) dx. \label{testint}
\ee
The case $\Delta =0$ describes either of the diagonal terms and is linearly
secular in time. This figure clearly shows the slow oscillations for
$t \gtrsim 1/\Delta$. \\

Therefore, the approximation (\ref{ltff}) is reliable in the long
time limit and upon  integration  with functions of compact support. We see that for large $t$, but $t\Delta \ll 1$,  $f_+(x,t,\Delta)f_-(x,t,\Delta) \rightarrow \pi t
\big[\delta(x-\Delta)+\delta(x+\Delta)\big]$ and for $\Delta
\rightarrow 0$, the product yields $2\pi t \delta(x)$. \\

The reduced density matrix (\ref{redrho}) allows us to obtain the time evolution of the
neutrino populations and coherences, namely
\be
n_i(\vp,t) = \mathrm{Tr}\, \rho_r(t) a^\dagger_i(\vp)a_i(\vp) ~~;~~\mathcal{C}_{ij}(\vp,t) =  \mathrm{Tr} \, \rho_r(t) a^\dagger_i(\vp)a_j(\vp)~,~i\neq j \label{rhoel}
\ee
where the annihilation and creation
operators are in the Schroedinger picture. In the long time limit and using the results above we find
\be
n_1(\vp,t) = t ~  \Gamma_1(\vp) \cos^2\theta ~~;~~  n_2(\vp,t) = t ~ \Gamma_2(\vp) \sin^2\theta, \label{nuprod}
\ee
where
\be \Gamma_{1,2}  = \frac{2\pi\,g^2}{8 E^W_{\vk}} \int
           \frac{d^3\vQ}{(2\pi)^3\,E^e_{\vQ}\,\Omega_{1,2}}
        \delta\big( E^W_{\vk} -E^e_{\vQ}-\Omega_{1,2}\big)\label{partialwidths}
\ee
are the partial widths for the decay of the $W$ into the charged lepton and the neutrino mass eigenstates. Similarly,
\bea
\mathcal{C}_{12}(\vp,t) = \mathcal{C}^\dagger_{21}(\vp,t) &=&
            \frac{2\pi\,g^2 \sin 2\theta}{32 E^W_{\vk}}~\frac{\sin[t\Delta]}{\Delta}~e^{-i\Delta t} \nn \\
            && \int
           \frac{d^3\vQ}{(2\pi)^3\,E^e_{\vQ}\,\sqrt{\Omega_{1}\,\Omega_{2}}}\Big[ \delta\big( E^W_{\vk} -E^e_{\vQ}-\Omega_{1}\big)+
            \delta\big( E^W_{\vk} -E^e_{\vQ}-\Omega_{ 2}\big) \Big]. \nn \\
           \label{coher}
\eea
$dn_i(\vp,t)/dt$ yields the \emph{production  rate} of the neutrino
           mass eigenstates from the decay of the $W$, and the coherences $\mathcal{C}_{ij}$ are non-vanishing at any finite time. In Ref. \cite{glashow}, these coherences vanish
           as a consequence of the product of delta functions on the different mass shells
           of the mass eigenstates. The coherences vanish in the formal infinite time
           limit because of the oscillatory behavior averages out on time scales $t \gg 1/\Delta$. However, we would like emphasize that on the experimental situation, the time
           scales involved (or rather length scales) are of order $1/\Delta$ as these are
           the scales on which oscillatory phenomena are revealed.
           Taking $\Omega_1 \sim \Omega_2 \sim \oO$ in the
           denominators in (\ref{partialwidths},\ref{coher}), it
           follows that \be \mathcal{C}_{12}(\vp,t) \simeq \frac{\sin 2\theta}{2}\frac{\sin[t\Delta]}{\Delta}~e^{-i\Delta t}
           ~\frac{1}{2}\big[\Gamma_1+\Gamma_2\big]. \label{cij}\ee
           Therefore, the coherences are of the same order of the
           population terms on time scales $t\leq 1/\Delta$, but
           average out for $t\gg 1/\Delta$. This clearly shows that
           coherence is maintained over the oscillation time scale.

\subsection{Disentangling the neutrino: a two-time measurement}

As we have discussed previously, a long baseline experiment is actually a \emph{two time
measurement}, as the charged lepton produced at the interaction vertex
at the \emph{source} is detected by the near detector. This
``measurement'' of the charged lepton disentangles the neutrinos from
the charged lepton in the quantum state
(\ref{finstate}) \cite{glashow}. The detection of the charged lepton
at the near detector projects the quantum state (\ref{finstate}) at
the observation time $t_S$ onto the single particle charged lepton
state $e^{-iE^e_{\vQ}t_S}~\Big|e_{\vQ}  \rangle$ disentangling the
neutrino states into the ``collapsed'' state
\bea
\big|
\mathcal{V}_{e}(\vP,t_S)\rangle & \equiv & \langle e_{\vQ}
\big|\Psi(t_S)\rangle ~e^{ iE^e_{\vQ}t_S}  \nn \\
&=& i\frac{g~e^{-iE_S
\frac{t_S}{2}}}{2\sqrt{2VE^W_{\vk} E^e_{\vQ}}}~
        \Bigg\{
    \frac{\sin\theta}{\sqrt{\Omega_{2, \vP}  }}~\big|\nu_{2,\vP}\rangle e^{ -i\Omega_{2,\vP} \frac{t_S}{2}}
    \left[\frac{\sin\Big[\big(E_S-\Omega_{2,\vP} \big)\frac{t_S}{2}\Big]}{    (E_S-\Omega_{2,\vP} )/2} \right]
    \nonumber \\ & & + \, \frac{\cos\theta}{\sqrt{ \Omega_{1,\vP} }}~\big|\nu_{1,\vP}\rangle
   e^{ -i\Omega_{1,\vP} \frac{t_S}{2}}
    \left[\frac{\sin\Big[\big(E_S-\Omega_{1,\vP} \big)\frac{t_S}{2}\Big]}{    (E_S-\Omega_{1,\vP} )/2} \right]
     \Bigg\}~; \label{colafinstate} \\
     && E_S = E^W_{\vk}-E^e_{\vQ}~;~\vP = \vk-\vQ \nn
\eea


We note that up to a phase the coefficient functions that multiply $\big|\nu_{1,2}\rangle$ are \be
\frac{\sin\Big[(E_S-\Omega_1)\frac{t_S}{2}
      \Big]}{(E_S-\Omega_1)/2} ~~,~~\frac{\sin\Big[(E_S-\Omega_2)\frac{t_S}{2}
      \Big]}{(E_S-\Omega_2)/2}, \label{senos}
\ee
respectively. In the limit of $t_S \rightarrow \infty$, these become $2\pi \delta(E_S-\Omega_1)$ and $2\pi \delta(E_S-\Omega_2)$, respectively. Therefore, in this limit, for a fixed $E_S$, one of the quantum states will be projected out. However, as we insist on keeping a \emph{finite}
time interval, we will keep $t_S$ finite. \\

The state $\big|\mathcal{V}_{e}(\vP,t_S)\rangle$ then evolves forward in time with the full Hamiltonian
\be
\big|\mathcal{V}_{e}(\vP, t)\rangle = e^{-iH_0 t} U(t,t_S) e^{iH_0 t_S} \big|\mathcal{V}_{e}(\vP,t_S)\rangle. \label{Nuoft}
\ee
The ``free'' evolution is obtained by setting  to lowest order $U(t,t_S)=1$, leading to
\bea
\big|\mathcal{V}_{e}(\vP, t)\rangle & = &  i\frac{g~e^{-iE_S \frac{t_S}{2}}}{2\sqrt{2VE^W_{\vk} E^e_{\vQ}}}~
        \Bigg\{
    \frac{\sin\theta}{\sqrt{\Omega_{2, \vP}  }}~\big|\nu_{2,\vP}\rangle
    \Bigg[\frac{ \sin\Big[\big(E_S-\Omega_{2,\vP} \big)\frac{t_S}{2}\Big] }{
    (E_S-\Omega_{2,\vP} )/2} \Bigg]~e^{-i\Omega_{2,\vP}\frac{t_S}{2}}~e^{-i\Omega_{2,\vP}(t-t_S)}
    \nonumber \\ & + & \frac{\cos\theta}{\sqrt{ \Omega_{1,\vP}}}~\big|\nu_{1,\vP}\rangle
    \Bigg[\frac{\sin\Big[\big(E_S-\Omega_{1,\vP} \big)\frac{t_S}{2}\Big] }{ (E_S-
    \Omega_{1,\vP} )/2} \Bigg] e^{-i\Omega_{1,\vP}\frac{t_S}{2}}~e^{-i\Omega_{1,\vP}(t-t_S)}
     \Bigg\}. \label{timev}
\eea
The  phase factors $e^{-i\Omega_j t_S/2}$ multiplying each mass eigenstate are the consequence of the phase build-up during
    the time evolution from the production vertex until the detection of the charged lepton
    and the collapse of the wave function. These can be absorbed into the definition of the states $|\nu_{1,2}\rangle$ at the fixed time $t_S$. \\

The expression (\ref{timev}) features the factors
\be
\frac{\sin\Big[\big(E_S-\Omega_{j} \big)\frac{t_S}{2}\Big] }{ (E_S-
    \Omega_{j } )/2}, \label{dels}
\ee
which as ${t_S\rightarrow \infty}$ becomes $2\pi\, \delta(E_S-\Omega_{j } )$. These factors, which are a direct consequence of the neutrino state being produced by the decay of the ``parent'' particle (here the W) into an \emph{entangled}   quantum state, distinguish Eq. (\ref{timev}) from the familiar quantum mechanical description. These factors emerge from the (approximate) energy conservation at the decay vertex.
Again, in the $t_S \rightarrow \infty$ limit, if the energy-momentum of the parent particle and the charged lepton are both certain,
\emph{only one} of the mass eigenstates will be produced but not both. However, writing $\Omega_{1,2}$ as in Eq. (\ref{URO}), it follows that for $t_S \ll 4\oO/\delta m^2$, the width of the ``diffraction'' functions is \emph{much larger} than the frequency difference $\Delta$ and there is a substantial overlap between these ``approximate'' delta functions. Only for $t_S \geq t_{osc} = 2\pi/\Delta$ are the two peaks at $E_S -\oO = \pm \Delta$ actually resolved, whereas for $t\ll t_{osc}$, the two peaks are unresolved, ``blurred'' into one broad peak at $\oO$. Thus, we can use the approximation
\be
\frac{\sin\Big[(E_S-\Omega_1)\frac{t_S}{2}
      \Big]}{(E_S-\Omega_1)/2} \simeq \frac{\sin\Big[(E_S-\Omega_2)\frac{t_S}{2}
      \Big]}{(E_S-\Omega_2)/2}\simeq \frac{\sin\Big[(E_S-\oO )\frac{t_S}{2}
      \Big]}{(E_S-\oO )/2},
      \label{equalsenos}
\ee
for $t_S \ll 2\pi/\Delta$. \\

To illustrate the validity of the above approximation, let us
consider the case in which the typical size of the near detector is
a few meters across. In a typical experiment, the charged lepton
emerging from the interaction vertex travels this distance within a
time scale $t_S \approx 10^{-8}$ s, leading to an energy uncertainty above  $\Delta E \sim \hbar/t_S \sim
10^{-7}\,\mathrm{eV}$. Taking as an example $\delta m^2 \sim
10^{-4}\,\mathrm{eV}^2$; $\oO \sim E_S \sim 100\,\mathrm{MeV}$, it
follows that $\delta m^2 /\oO \sim 10^{-12}\,\mathrm{eV} \ll \Delta
E$. Therefore, the detection at the near detector cannot resolve the
energy difference between the mass eigenstates and the approximation
(\ref{equalsenos}) is justified. \\

Another approximation we can use in (\ref{colafinstate}) is $\Omega_1 \simeq \Omega_2\simeq \oO$,  for $\oO \gg \Delta$. Absorbing the phase factors $e^{-i\Omega_j t_S/2}$
       into the definition of the states $|\nu_j\rangle$,   the time evolved disentangled state is then approximately given by
\be
\big|\mathcal{V}_{e}(\vP,t)\rangle \simeq \frac{i g }{ \big[8
VE^W_{\vk} E^e_{\vQ}\oO\big]^\frac{1}{2}}
\left[\frac{\sin\Big[\big(E_S-\oO \big)\frac{t_S}{2}\Big]}{
(E_S-\oO  )/2} \right]~\Bigg\{     {\sin\theta} ~\big|\nu_{2,\vP}\rangle~e^{-i\Omega_{2,\vP}(t-t_S)}
     + \, {\cos\theta} ~\big|\nu_{1,\vP}\rangle~e^{-i\Omega_{1,\vP}(t-t_S)}
    \Bigg\},  \label{colafinstatesmallt}
\ee
for $t_S \ll 2\pi/\Delta$.

The state inside the brackets is identified as the usual quantum mechanical state that is
   time evolved from the ``electron'' neutrino state, which is prepared initially at $t_S$.
    From this analysis, we see that there are two conditions required for the disentangled neutrino
     state to be identified with the familiar quantum mechanical state. The two conditions are $\delta m^2/\oO^2 \ll1$ and $t_S \ll t_{osc} \sim 2\pi \oO/\delta m^2$. The former is always satisfied for neutrinos with
     $\delta m^2 \sim 10^{-3}-10^{-4} \,\mathrm{eV}^2,\ \oO > \,\mathrm{few~MeV}$, while
      the latter is fulfilled for near-detection of the charged lepton
       in long-baseline experiments.
       The latter condition implies that
       the neutrino state is \emph{disentangled} before oscillations can
       occur. In a long-baseline experiment this is achieved if the charged lepton, which is entangled with the neutrino at the production
        vertex, is measured at the near detector .

\subsection{Transition amplitudes and event rates}

The number of charged lepton events with momentum $\vQ$ at the near
detector, at time $t_S$ is given by
\be
n_e(\vQ,t_S)= \langle
\Psi_e(t_S)| a^\dagger_{e}(\vQ) a_e(\vQ) | \Psi_e(t_S) \rangle
=\langle \mathcal{V}_{e}(\vP,t_S) \big|
\mathcal{V}_{e}(\vP,t_S)\rangle.  \label{neND}
\ee
For $t_S \ll t_{osc}=2\pi/\Delta$ and $\oO \gg \Delta$, using the approximations
leading  to (\ref{colafinstatesmallt}), we obtain the differential
detection rate at the \emph{near detector}
\bea
(2\pi)^3\, \frac{dN^{\rm (ND)}_e}{d^4x\, d^3\vQ}&=&\frac{dn_e(\vQ,t_S)}{dt_S} \nn \\
&=&  \frac{2 g^2 }{ 8 V E^W_{\vk} E^e_{\vQ}\, \oO  } \,\frac{\sin\Big[\big(E_S-\oO
\big) {t_S }\Big] }{    (E_S-\oO  ) }  \nn \\
 &\simeq&  \frac{ 2\pi g^2 }{ 8
VE^W_{\vk} E^e_{\vQ }\,\oO  }~ \delta(E_S-\oO ), \label{diffrateND}
\eea
where at the last step we have replaced the diffraction function by the delta function.  This can be justified as follows. For $t_S \sim 10^{-8} \,\mathrm{s}$, the width of this function (the resolution)  in energy is $\sim 10^{-7}\,\mathrm{eV}$. Since the typical energy in a long-baseline experiment is $ \gtrsim 40-100 \,\mathrm{MeV}$, the error incurred in replacing the diffraction function by a delta function is smaller than one part in $10^{15}$. \\


 We can also obtain the transition amplitude for the disentangled  state to produce
     a final charged lepton and another $W$ particle at the far detector at time $t_D$,
     where $t_D-t_S \sim L$ and $L$ is the baseline. It is given by
\bea
\mathcal{A}_{\alpha \rightarrow \beta} &=& \langle W(\vk_D),l_\beta(\vp_D)\big|
     \,e^{-iH(t_D-t_S)}\,\big|\mathcal{V}_{e}(\vP,t_S)\rangle  \nn \\
     &=& e^{-iE_D t_D}\, \langle W(\vk_D),l_\beta(\vp_D)\big|\,U(t_D,t_S)\,e^{iH_0t_S}\,\big|\mathcal{V}_{e}(\vP,t_S)\rangle. \label{Aabcola}
\eea
The disappearance and appearance amplitudes are then given by
\bea
\mathcal{A}_{e\rightarrow e} & = & -g^2\,\Pi\,(2\pi)^3 \delta(\vk_S-\vp_S-\vk_D-\vp_D)  \times \nn \\  & & \Bigg\{ \frac{\cos^2\theta}{2\Omega_{1,\vP}} e^{-i\Omega_{1,\vP}\frac{t_D}{2}}\, \Bigg[\frac{\sin\Big[\big(E_S-\Omega_{1,\vP} \big)\frac{t_S}{2}\Big] }{ (E_S-
    \Omega_{1,\vP} )/2} \Bigg] \, \Bigg[\frac{\sin\Big[\big(E_D-\Omega_{1,\vP} \big)\big(t_D-t_S\big)/2\Big] }{ (E_D-\Omega_{1,\vP} )/2} \Bigg]
    \nonumber \\ &  & ~~+ \frac{\sin^2\theta}{2\Omega_{2,\vP}} e^{-i\Omega_{2,\vP} \frac{t_D}{2}}\,
     \Bigg[\frac{\sin\Big[\big(E_S-\Omega_{2,\vP} \big)\frac{t_S}{2}\Big] }{ (E_S-
    \Omega_{2,\vP} )/2} \Bigg] \, \Bigg[\frac{\sin\Big[\big(E_D-\Omega_{2,\vP} \big)\big(t_D-t_S\big)/2\Big] }{ (E_D-
    \Omega_{2,\vP} )/2} \Bigg]\Bigg\} \nn \\
    \label{disadisent}
\eea
and
\bea
\mathcal{A}_{e\rightarrow \mu} & = & -g^2\,\Pi\,(2\pi)^3 \delta(\vk_S-\vp_S-\vk_D-\vp_D) \,\frac{\sin2\theta}{2} \times \nn \\
     && \Bigg\{ \frac{e^{-i\Omega_{1,\vP}\frac{t_D}{2}}}{2\Omega_{1,\vP}}  \, \Bigg[\frac{\sin\Big[\big(E_S-\Omega_{1,\vP} \big)
     \frac{t_S}{2}\Big] }{ (E_S-\Omega_{1,\vP} )/2} \Bigg] \,
      \Bigg[\frac{\sin\Big[\big(E_D-\Omega_{1,\vP} \big)\big(t_D-t_S\big)/2\Big] }{ (E_D-\Omega_{1,\vP} )/2} \Bigg]
    \nonumber \\ &   & ~~  - \frac{e^{-i\Omega_{2,\vP} \frac{t_D}{2}}}{2\Omega_{2,\vP}}
      \, \Bigg[\frac{\sin\Big[\big(E_S-\Omega_{2,\vP} \big)\frac{t_S}{2}\Big] }{ (E_S-
    \Omega_{2,\vP} )/2} \Bigg] \,
     \Bigg[\frac{\sin\Big[\big(E_D-\Omega_{2,\vP} \big)\big(t_D-t_S\big)/2\Big] }{ (E_D-
    \Omega_{2,\vP} )/2} \Bigg]\Bigg\},\nn \\
    \label{aperadisent}
\eea
with $\vP=\vk_S-\vp_S$.  In these expressions, we have used the same notation as in section (\ref{sec:3}), where $\Pi$ is given by (\ref{Pifactor}).
Implementing the same approximations leading to the factorized state
(\ref{colafinstatesmallt}), namely $\oO \gg \Delta$ and $\,t_S\Delta \ll
1$, we find the disappearance and appearance probabilities
\bea
\mathcal{P}_{e\rightarrow e} (t_D) & = &  \Big(\frac{g^2 \Pi}{2\oO_{\vP}}\Big)^2 (2\pi)^3 \, V\,\delta(\vk_S-\vp_S-\vk_D-\vp_D)\,   2\pi\,t_S\,\delta(E_S-\oO_{\vP}) \nonumber \\ && \Bigg\{\cos^4\theta \, f^2_+(x,t,\Delta) + \sin^4\theta \,f^2_-(x,t,\Delta)+ 2\cos^2\theta \sin^2\theta \cos(t\Delta) f_+(x,t,\Delta)f_-(x,t,\Delta) \Bigg\}, \label{probee}\nonumber \\
 \mathcal{P}_{e\rightarrow \mu}(t_D)  & = &  \Big(\frac{g^2 \Pi}{2\oO_{\vP}}\Big)^2 (2\pi)^3 \, V\,\delta(\vk_S-\vp_S-\vk_D-\vp_D)\,   2\pi\,t_S\,\delta(E_S-\oO_{\vP})\, \frac{\sin^22\theta}{4}\nonumber \\ && \Bigg\{f^2_+(x,t,\Delta)+f^2_-(x,t,\Delta)-2\cos(t\Delta) f_+(x,t,\Delta)f_-(x,t,\Delta) \Bigg\}, \label{probemu}
\eea
where $t= t_D-t_S$ and $x=E_D-\oO_{\vP}$. Here,
\be
E_S = E^W_{\vk_S}-E^e_{\vp_S},~~ E_D= E^W_{\vk_D}+E^l_{\vp_D}\label{defs2},
\ee
and $f_{\pm}$ are given by Eq. \ref{fpm}. \\

In the long time limit, using $f_{\pm}(x,t,\Delta) \rightarrow 2\pi \,t\,\delta(x\pm\Delta)$ and (\ref{ltff}), we find
\bea
\mathcal{P}_{e\rightarrow e}(t_D)  & = &  \Big(\frac{g^2 \Pi}{2\oO_{\vP}}\Big)^2 (2\pi)^5 \, V\,\delta(\vk_S-\vp_S-\vk_D-\vp_D)\,   \,t_S\,\delta(E_S-\oO_{\vP}) \nonumber \\ && \Bigg\{\cos^4\theta \, t\,\delta(x+\Delta) + \sin^4\theta \,t\,\delta(x-\Delta)+ 2\cos^2\theta \sin^2\theta \,\frac{\sin(2t\Delta)}{2\Delta}\,\frac{1}{2}\Big[\delta(x+\Delta)+\delta(x-\Delta)\Big] \Bigg\}, \label{probeelt}\nonumber \\
 \mathcal{P}_{e\rightarrow \mu}(t_D)  & = &  \Big(\frac{g^2 \Pi}{2\oO_{\vP}}\Big)^2 (2\pi)^5 \, V\,\delta(\vk_S-\vp_S-\vk_D-\vp_D)\,   \,t_S\,\delta(E_S-\oO_{\vP})\, \frac{\sin^22\theta}{4}\nonumber \\ && \Bigg\{t\,\delta(x+\Delta)+t\,\delta(x-\Delta)-2\,\frac{\sin(2t\Delta)}{2\Delta}\,\frac{1}{2} \Big[\delta(x+\Delta)+\delta(x-\Delta)\Big]  \Bigg\}. \label{probemult}
 \eea

 These transition probabilities are \emph{very different} from the ones obtained in section (\ref{sec:probs})
  and from those obtained from the S-matrix approach. They feature the \emph{two} time scales $t_S$ and $t=t_D-t_S$ associated with the measurements at the near and far detector. They also feature energy conserving delta functions associated with the different mass eigenstates. \\

 There is a further simplification when $\oO \gg \Delta$. In this regime, when the probabilities (\ref{probeelt},\ref{probemult}) are \emph{integrated  over a smooth density of states}, the delta functions corresponding to the mass eigenstates yield the density of states at values $E_D = \oO \mp \Delta$. In typical experiments, where $\oO \sim 100 \,\mathrm{MeV}$ and $\delta m^2 \sim 10^{-3}\,\mathrm{eV}^2$, the density of final states must vary dramatically
 near $\oO$ to resolve the small interval $\Delta$, with $\Delta/\oO \lesssim 10^{-19}$. Therefore,
  understanding the probabilities as being integrated with a smooth final density of states insensitive to the mass difference,
   we can approximate $\delta(x\pm \Delta) \simeq  \delta(x)$. In this case, we can approximate the above expressions by
\bea
\mathcal{P}_{e\rightarrow e} (t_D) & = &  \Big(\frac{g^2 \Pi}{2\oO_{\vP}}\Big)^2 (2\pi)^5 \, V\,\delta(\vk_S-\vp_S-\vk_D-\vp_D)\,
   \,t_S\,\delta(E_S-\oO_{\vP})\,\delta(E_D-\oO_{\vP}) \nonumber \\ && \Bigg\{t\Big[\cos^4\theta   + \sin^4\theta \Big]+
   2\cos^2\theta \sin^2\theta \,\frac{\sin(2t\Delta)}{2\Delta}  \Bigg\}, \label{probeelt2} \\
 \mathcal{P}_{e\rightarrow \mu} (t_D) & = &  \Big(\frac{g^2 \Pi}{2\oO_{\vP}}\Big)^2 (2\pi)^5 \, V\,\delta(\vk_S-\vp_S-\vk_D-\vp_D)\,
   \,t_S\,\delta(E_S-\oO_{\vP})\,\delta(E_D-\oO_{\vP})\, \frac{\sin^22\theta}{2}\nonumber \\ && \Bigg\{t - \,\frac{\sin(2t\Delta)}{2\Delta} \Bigg\}.
    \label{probemult2}
\eea
The product $ \delta(E_S-\oO_{\vP})\,\delta(E_D-\oO_{\vP}) $ is an \emph{approximate} energy conservation at both production and detection vertices, where we have neglected $\Delta$, which is twice the energy difference between the mass eigenstates. However, even under these justified approximations, the probabilities (\ref{probeelt2},\ref{probemult2}) are very different from those obtained by the S-matrix approach, even after including the finite time corrections discussed in section (\ref{sec:probs}). They also differ greatly from the transition probabilities from the simple quantum mechanical argument.\\

Further insight can be gained by obtaining the phase space distribution of the number of charged leptons $l=e,\mu$ at the far detector
\be
(2\pi)^3\,\frac{dN_l}{d^3x \,d^3\vp_D}=n_l(\vp_D,t_D) = \langle \mathcal{V}_e(\vQ,t_D,t_S)\big|a^\dagger_l(\vp_D)a_l(\vp_D) \big| \mathcal{V}_e(\vQ,t_D,t_S)\rangle. \label{fardetn}
\ee
Here,
\be
\big| \mathcal{V}_e(\vQ,t_D,t_S)\rangle = e^{-iH_0 t_D}U(t_D,t_S)e^{iH_0t_S} \big| \mathcal{V}_e(\vQ,t_S)\rangle \label{tevolnu}
\ee
is the neutrino state disentangled at $t_S$ at the  near detector and time-evolved until it is detected at the far detector at time $t_D$. Not surprisingly, since up to order $g^2$,
the time evolved state contains a single lepton Fock state, we find that
\be
(2\pi)^3\,\frac{dN_e}{d^3x \, d^3\vp_D} = \mathcal{P}_{e\rightarrow e} (t_D) ~~;~~ (2\pi)^3\, \frac{dN_{\mu}}{d^3x \,d^3\vp_D} = \mathcal{P}_{e\rightarrow \mu} (t_D), \label{nemuFD}
\ee
with the probabilities $\mathcal{P}_{e\rightarrow e} (t_D)$ and $\mathcal{P}_{e\rightarrow \mu} (t_D)$ are given by (\ref{probeelt2},\ref{probemult2}).\\

Taking the \emph{time derivative} with respect to $t_D$, we obtain the differential charged leptons event \emph{rates} at the far detector
\bea
(2\pi)^3\frac{dN^{FD}_e}{d^3x \,dt \,d^3\vp_D}  & = &   \Big(\frac{g^2 \Pi}{2\oO_{\vP}}\Big)^2 (2\pi)^5 \, V\,\delta(\vk_S-\vp_S-\vk_D-\vp_D)\,   \,t_S\,\delta(E_S-\oO_{\vP})\,\delta(E_D-\oO_{\vP}) \nonumber \\ && \Bigg\{ \cos^4\theta   + \sin^4\theta  + 2\cos^2\theta \sin^2\theta \, \cos(2t\Delta)  \Bigg\} ,\label{FDee}
\eea
\bea
(2\pi)^3\frac{dN^{FD}_{\mu}}{d^3x \,dt\, d^3\vp_D} & = &
 \Big(\frac{g^2 \Pi}{2\oO_{\vP}}\Big)^2 (2\pi)^5 \, V\,\delta(\vk_S-\vp_S-\vk_D-\vp_D)\,
   \,t_S\,\delta(E_S-\oO_{\vP})\,\delta(E_D-\oO_{\vP})\, \frac{\sin^22\theta}{2}\nonumber \\ &&
   \Bigg\{1 -  \cos(2t\Delta) \Bigg\}. \label{FDemu}
\eea
Remarkably, these rates can be simply written as
\be
(2\pi)^3\frac{dN^{FD}_\beta}{d^3x \,dt\, d^3\vp_D} =
   (2\pi)^3\frac{dN^{ND}_\alpha}{d^3x \,d^3\vp_s}\,P_{\alpha \rightarrow \beta}(t) \,d\Gamma_{\nu_\beta \rightarrow W\,l_\beta}, \label{factorized}
\ee
where we have used the expression (\ref{diffrateND}) for the differential charged lepton event rate at the near detector, integrated in $t_S$,
\be
d\Gamma_{\nu_\beta \rightarrow W\,l^\beta} = \frac{(2\pi)^4\,g^2\,V}{8V^3E^W_{\vk_D}E^{l^\beta}_{\vp_D}\oO}
     \, \delta(\vk_S-\vp_S-\vk_D-\vp_D)\,\delta(E_D-\oO_{\vP}) \label{prodrateFD}
\ee
is the charged lepton production rate from the reaction $\nu_\beta \rightarrow W\,l_{\beta}$ for a \emph{flavor} neutrino of energy $\oO$ and $P_{\alpha \rightarrow \beta}(t)$ are the disappearance ($\alpha = \beta$) or appearance ($\alpha \ne \beta$) \emph{quantum mechanical} transition probabilities.

The remarkable aspect of the final result (\ref{prodrateFD}) is the \emph{factorization}
 of the different processes contributing to the far detector event \emph{rate}, namely the number
 of events at the near detector multiplies the quantum mechanical transition probability which in
 turn multiplies the production rate at the vertex in the far detector. This factorization is a
 distinct consequence of the \emph{two time analysis}, of the disentanglement of the neutrino at
  the near detector, along with the approximations invoked in the resolution of the energy
  conserving delta functions. We emphasize that this factorization in terms of the
 quantum mechanical transition probabilities \emph{only} applies to the detection \emph{rate}
  defined by taking the time derivative,  but \emph{not} to the total number of events or to the
 rate defined by simply \emph{dividing} by the time scale.

\section{A model for the GSI anomaly}
Recent experiments at the Experimental Storage Ring (ESR) at
     GSI in Darmstadt have revealed an unexpected time dependent
     modulation in the population of daughter ions
     ${}^{140}Ce^{58+}$ from the electron capture decay ${}^{140}Pr^{58+}\rightarrow
     {}^{140}Ce^{58+}+\nu_e$ \cite{gsi1,gsi2}, a phenomenon that has been dubbed the ``GSI
     anomaly.'' There have been some works that try to explain this remarkable time dependent rate of the parent   ion decay as an
     interference between the neutrino mass eigenstates in the decay
     reaction \cite{gsi1,gsi2,kienle,ivanov1,ivanov2,ivanov3,faber,lipkingsi}. However, this interpretation has been
     re-examined and criticized \cite{giuntygsi1,giuntygsi2,burka,kienert,peshkin}. \\

    The authors in Refs. \cite{kienle,ivanov1,ivanov2,ivanov3,faber} obtain the decay rate of the parent ion
     by adding coherently the amplitudes and then obtaining the probabilities, in which case the modulation arises from the interference of the mass eigenstates in the squaring of the
     amplitudes. The authors in Refs. \cite{giuntygsi1,giuntygsi2,burka,kienert} criticize this approach, stating that it is not the amplitudes that must be summed coherently but the probabilities, corresponding to an \emph{incoherent} addition of the contributions from the
     mass eigenstates. This approach does not lead to any modulation as the amplitudes for the different mass eigenstates do not interfere.\\

     Rather than focusing on either one of these approaches, we analyze the situation \emph{differently}, by obtaining \textbf{the time evolution of the population of the parent and daughter particles}. In this approach, we recognize directly the decay
     rate of the parent particles (or production rate of the daughter) without the necessity to invoke a coherent sum over amplitudes or a sum over probabilities. \\

     We   study the time evolution of the populations by modeling the
     situation in terms of the decay of a parent particle via charged current interactions
      into a  charged lepton and its associated neutrino. The interaction vertex is given by $W\,e\,\nu_e$, where we simply take the parent to be the  $W$ and the daughter to be the $e$-charged lepton. \\

      Let us consider an initial parent particle state $\big|W(\vk) \rangle$ that is prepared  at time $t=0$. The evolution of the number of parent and daughter particles is obtained from
\bea
N_W(\vk,t) &=& \langle W(\vk)\big|\, e^{iHt}~ a^\dagger_W(\vk) a_W(\vk) ~ e^{-iHt}\,\big|W(\vk)\rangle, \nn \\
       n_e(\vQ,t) &=& \langle W(\vk)\big|\,e^{iHt}~ a^\dagger_{e}(\vQ) a_e(\vQ) ~ e^{-iHt}\,\big|W(\vk)\rangle, \label{numbers}
\eea
where the annihilation and creation operators are in the Schroedinger picture. We note that $e^{-iHt} = e^{-iH_0t}U(t,0)$ and that the number operators commute with the free field Hamiltonian.  We also note that $U(t,0)\big|W\rangle = \big|W\rangle + \big|\Psi_e(t)\rangle^{(1)}+\big|\Psi_e(t)\rangle^{(2)}+\cdots$
where
\be \big|\Psi_e(t)\rangle^{(1)} =  ig \int_0^t dt_1 \int d^3x_1 \Big[W(\vec{x}_1,t_1) e(\vec{x}_1,t_1)\nu_e (\vec{x}_1,t_1)\big)\Big] \big|W(\vk)\rangle, \label{psi1}
\ee
and
\bea
\big|\Psi_e(t)\rangle^{(2)} = - g^2 \int_0^t dt_1 \int d^3x_1 \int_0^{t_1} dt_2 \int d^3x_2 \Big[W(\vec{x}_1,t_1) e(\vec{x}_1,t_1)\nu_e (\vec{x}_1,t_1)\big)\Big]\nn \\ \Big[W(\vec{x}_2,t_2) e(\vec{x}_2,t_2)\nu_e (\vec{x}_2,t_2)\big)\Big] \big|W(\vk)\rangle \,,\label{psi2}
\eea
with $\nu_e = \cos\theta\, \nu_1 +\sin\theta\, \nu_2 $. Since the (entangled) state
$\big|\Psi_e(t)\rangle^{(1)}$ has one daughter particle (electron) and the initial
state has none, it is clear that to lowest order, the number of daughter particles is
\be
n_e(\vQ,t) = {}^{(1)}\langle \Psi_e(t)\big| a^\dagger_e(\vQ) a_e (\vQ) \big|\Psi_e(t)\rangle^{(1)} =
       \langle \mathcal{V}_{e}(\vQ,t ) \big|\mathcal{V}_{e}(\vQ,t )\rangle, \label{enumero}
\ee
where $\big|\mathcal{V}_{e}(\vQ,t )\rangle$ is the ``collapsed'' state (\ref{colafinstate}). We find the production rate of the daughter particle
\be
 \frac{dn_e(\vQ,t)}{dt} = \frac{g^2}{8V\,E^W_{\vk} E^e_{\vQ}}\Bigg[ \frac{\cos^2\theta}{\Omega_1} \,
        \frac{2\sin\big[( E^W_{\vk} -E^e_{\vQ}-\Omega_1)t\big]}{( E^W_{\vk} -E^e_{\vQ}-\Omega_1)}+ \frac{\sin^2\theta}{\Omega_2} \,
       \frac{2\sin\big[( E^W_{\vk} -E^e_{\vQ}-\Omega_2)t\big]}{( E^W_{\vk} -E^e_{\vQ}-\Omega_2)}\Bigg].  \label{prodaughter}
\ee

If at this stage we take the long time limit, using the identity
\be
 \frac{2\sin\big[( E^W_{\vk} -E^e_{\vQ}-\Omega_{j})
      t \big]}{\big( E^W_{\vk} -E^e_{\vQ}-\Omega_{j}  \big)}
           \rightarrow 2\pi  \delta\big( E^W_{\vk} -E^e_{\vQ}-\Omega_{j}
      \big), \label{delts}
\ee
we find the total number of daughter particles produced as a function of time
       $n_e(t) = \sum_{\vQ} n_e(\vQ,t)$ is given by
\be
n_e(t) = \Big[\Gamma_1\,
            \cos^2\theta   + \Gamma_2\,\sin^2\theta  \Big]\,t, \label{neoftinfP}
\ee
where
          $\Gamma_{1,2}$ are the partial widths given by
          (\ref{partialwidths}). In the rate (\ref{prodaughter}),
          there is \emph{no interference} between the mass
          eigenstates since $|\nu_{1,2}\rangle$ are orthogonal.  It is straightforward to
            confirm that
\be
n_e(t) = {}^{(1)}\langle \Psi_e(t)\big|  \Psi_e(t)\rangle^{(1)} = \sum_{\vQ} \langle \mathcal{V}_{e}(\vQ,t ) \big|\mathcal{V}_{e}(\vQ,t )\rangle \,, \label{nid}
\ee
a result that will play an important role below.\\

\begin{figure}[h!]
\begin{center}
\includegraphics[height=3in,width=3in,keepaspectratio=true]{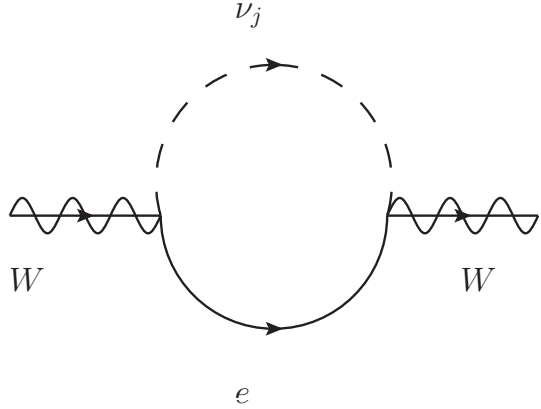}
\caption{Self-energy of the parent particle $W$ } \label{fig:se}
\end{center}
\end{figure}

            The calculation of the parent population is slightly more involved. Since the first order state does not have
any parent particle $W$, we must consider the second order state (\ref{psi2}).
 To second order, there are several contributions, but the only one that is relevant is the process
 in which the first vertex at $(\vec{x_2},t_2)$ annihilates the initial $W$ creating the intermediate state with one $(e,\nu_e)$
 entangled pair, while the second interaction vertex  at $(\vec{x}_1,t_1)$   \emph{annihilates} this  $(e,\nu)$ pair in the intermediate state
 and creates the $W$, which has non-vanishing overlap with $\big|W\rangle$. This process is depicted in Fig. (\ref{fig:se}) and is recognized
 as the self-energy of the parent particle. \\

 Thus to lowest order in $g$, \be N_W(\vk,t) = 1+2\, \mathrm{Re}\Big[\langle W \big|\Psi_e(t)\rangle^{(2)} \Big].  \label{nWex}\ee
  It proves more convenient to calculate $dN_W(\vk,t)/dt$,
    for which we find\footnote{Effectively, we are obtaining the Boltzmann equation for the parent particle,
  neglecting the build-up of the population.}
\be
\frac{dN_W(\vk,t)}{dt}= -\frac{2 \,g^2}{8 E^W_{\vk}} \int
           \frac{d^3\vQ}{(2\pi)^3\,E^e_{\vQ}} \Bigg\{ \frac{\cos^2\theta}{\Omega_{1}}~
         \frac{ \sin\Big[\big( E^W_{\vk} -E^e_{\vQ}-\Omega_{1}\big)t\Big]}{\big( E^W_{\vk} -E^e_{\vQ}-\Omega_{1}\big)} +\frac{\sin^2\theta}{\Omega_{2}}~
         \frac{ \sin\Big[\big( E^W_{\vk} -E^e_{\vQ}-\Omega_{2}\big)t\Big]}{\big( E^W_{\vk} -E^e_{\vQ}-\Omega_{2}\big)} \Bigg\}.  \label{dnedt}\ee  In the long time limit this becomes \be \frac{dN_W(\vk,t)}{dt} = -\Big[\Gamma_1\,
            \cos^2\theta   + \Gamma_2\,\sin^2\theta  \Big]. \label{Wrate}
\ee
Clearly, $dN_W(\vk,t)/dt = - dn_e(t)/dt$ as the decay of the parent population results in the growth of the daughter population with the same rate. This is a consequence of unitarity and we can see this by substituting (\ref{nid},\ref{nWex})  into the unitarity condition
\be
\langle W(\vk)\big|U^\dagger(t,0)U(t,0)\big|W(\vk)\rangle = 1 = 1+ {}^{(1)}\langle \Psi_e(t)\big|  \Psi_e(t)\rangle^{(1)}+ 2\, \mathrm{Re}\Big[\langle W \big|\Psi_e(t)\rangle^{(2)}\rangle \Big] +\mathcal{O}(g^3)+\cdots \label{unitarity}
\ee

The interpretation of (\ref{Wrate}) in terms of the Feynman diagram (\ref{fig:se}) also makes the conclusion of lack of interference manifest.
 The decay rate is the imaginary part of the self-energy. Since the correct propagating degrees of freedom are the neutrino \emph{mass eigenstates},
  the self-energy is the \emph{sum} of the self-energy diagrams with the neutrino mass eigenstates propagating inside the loop. Therefore, it simply follows that the total decay width is the \emph{sum} of the partial
decay widths on the mass eigenstates without interference. The real time calculation presented above confirms this result directly from the evolution of the parent and daughter populations.
Thus, we confirm the analysis of Refs. \cite{giunti1,giunti2,giunti3,giunti4,giunti5,giunticohe,burka,peshkin} that there is \emph{no interference} of mass eigenstates and we conclude that the GSI anomaly \emph{cannot} be explained in terms of the interference of mass eigenstates in the decay. \\

\section{Conclusions and Discussions}
In this article, we carry out an in-depth review on our recent discovery on the theory of neutrino oscillations.
we study the dynamics of mixing and oscillations in quantum field theory
directly in real and finite time. The setting is a bosonic model that reliably describes
charged current-weak interactions. This allows us to extract the relevant aspects without the
peripheral complications associated with spinors. We begin by obtaining the in-out transition
amplitudes and probabilities in a long but finite time interval for appearance and disappearance processes and compare these to the S-matrix results. To illustrate the effects of the finite time interval, we consider the simplest setting with plane wave states, discussing how a wave-packet treatment needs to be modified to include the finite time contributions. We find finite time contributions that display the oscillatory behavior resulting from the interference of mass eigenstates and show that these corrections can be of the same order of or larger than the S-matrix result for \emph{appearance} experiments. A deeper analysis of the different contributions lead us to argue that the in-out treatment is ill-suited to describe long-baseline experiments. The main reason is that in the production vertex at the source or near detector, the neutrino is produced in an \emph{entangled state} with the charged lepton. \\

The concept of the finite time interval is compatible with the recent observation that the neutrino forms an \emph{entangled state} with the charged lepton produced at the near detector \cite{giunticohe,glashow,losecco1,losecco2} and enable us to study the details of the dynamics of this entanglement. If the charged lepton (or daughter particle) produced at the near detector is not measured, tracing out this degree of freedom yields a density matrix for the neutrino. The off-diagonal density matrix elements in the mass basis are a measure of coherence and show distinctly the oscillatory interference effects. Diagonal and off diagonal matrix elements are of the same order of magnitude during time scales $t \lesssim t_{osc} \simeq 2 E/\delta m^2$, with $E$ the typical neutrino energy, revealing that coherence survives during the time scale $t_{osc}$. On the other hand, the measurement of the charged lepton at the near detector \emph{disentangles} the neutrino state, and it is the further time evolution of this disentangled state with the interaction Hamiltonian that leads to the production of charged leptons at the far detector. Thus, the process of production and detection in long-baseline experiments involves \emph{two different time scales}: the measurement of the charged lepton at the near detector determines the first time scale at which the neutrino state is disentangled, while the measurement of the charged lepton at the far detector is the second and longer time scale. \\

The form of the amplitudes (\ref{Aee2},\ref{Aemu2}) and probabilities (\ref{Pee},\ref{Pemu})  obtained from the in-out formulation are fundamentally different from those obtained from the
time evolution of the disentangled state (\ref{disadisent}, \ref{aperadisent}). We can see why this is so by noticing the difference in how and when the interaction vertices act on the state. In the transition amplitudes (\ref{Aee2},\ref{Aemu2}), the second order matrix element
involves the time ordered product or alternatively the nested time integrals (see \ref{2ndord} ), which result in overall energy conservation. On the other hand, in the amplitudes obtained from the disentangled state, the perturbation acts at different and non-overlapping times. A first order vertex creates and evolves the entangled state until the measurement of the charged lepton disentangles the neutrino state, while \emph{another} first order vertex propagates the
disentangled neutrino and creates the final charged lepton that is measured at the far detector. There is another way to understand this. As made explicit by time dependent perturbation theory (see subsection \ref{sec:look}), there are \emph{two} contributions (corresponding to the time ordering) whose sum yields the in-out amplitudes. These contributions are also manifest in the in-out probabilities. Whereas in the second case,
there is only one contribution (as discussed in subsection \ref{sec:look}). These differences explain the fundamentally different form between the amplitudes and probabilities obtained
from the in-out formulation, even at finite time, and those obtained from the disentanglement at the near detector and further evolution of the neutrino that leads to the charged lepton at the far detector. \\

For long baseline neutrino oscillation experiments corresponding to the case of the disentangling time much smaller than the baseline, We obtain the charged lepton event \emph{rate} at the far detector and factorize this rate into a product of the number of charged lepton events per phase space at the near detector times the \emph{quantum mechanical} transition probability, times the differential production rate at the far detector. This factorization is a direct consequence of the \emph{disentanglement} of the neutrino state at the near detector along with the approximation that the density of final states is a smooth function and for large energies, it is not sensitive to the energy difference of the mass eigenstates. This factorization was \emph{assumed} in Ref. \cite{akmerev} and physical arguments were proposed for its validity in Ref. \cite{glashow}. However, in this article, we show that the time evolution of the disentangled neutrino state unambiguously leads to the factorization, under the approximations discussed above. We would like to emphasize that the factorization in terms of the quantum mechanical probabilities is only valid for the event \emph{rate} at the far detector. It is \emph{not} valid for the total number of events. This latter quantity is also factorizable, but not in terms of the transition probability, but its time integrated version, which leads to a \emph{different energy dependence of the oscillatory terms} as can be gleaned from the second lines in Eqs. (\ref{probeelt2},\ref{probemult2}). \\

More interestingly, the appearance and disappearance amplitudes (\ref{disadisent}, \ref{aperadisent}) depend on both the disentangling time scale and the final detection time scale. For short baseline neutrino oscillation experiments, e.g., LSND experiment \cite{Athanassopoulos1,Athanassopoulos2,Athanassopoulos3,Athanassopoulos4}, the disentangling time can be of the same order as the detection time, which is also of the same order as the oscillation time scale. In this case, the disentangling time introduces an \emph{extra modulation with energy} (see (\ref{disadisent}, \ref{aperadisent})) which may yield phenomenologically interesting modifications in the interpretation and analysis of data. Based on the dynamics of entanglement and disentanglement discussed here, a recent study by Boyanovsky \cite{Boyanovsky:2011xq} analyzes the influence of the disentanglement on neutrino oscillation formulas of short baseline neutrino oscillation experiments by introducing yet another time scale describing the decay length of the parent particle. It was found that both the disentanglement length scale and the source lifetime lead to a suppression of the oscillation probabilities in short baseline experiments, and imply that fits to the experimental data based on the standard quantum mechanical formulas underestimate both the mixing angle and the mass difference between different generations of neutrinos.\\

Although our study has been carried out in terms of plane waves, we can extrapolate our results to include the case of wave packets. Let us consider the initial state in (\ref{newstate}).   Instead of being the plane wave Fock state $|W(\vk)\rangle$, let it be the localized state       $|\widetilde{W}(S)\rangle$ given by (\ref{locstates}). Let us also assume that the charged lepton measured at the near detector is measured with a wave-function represented by a
wave-packet (\ref{locstates}) $|\widetilde{l}_e(S)\rangle$. The neutrino is therefore
disentangled into a wave-packet state, obtained from (\ref{colafinstate}) by the convolution with the wave-packets of the $W$ and $e$. This resulting state will be described by a wave packet whose localization length is determined by the production and disentangling processes. Upon further time evolution, this wave packet will spread and split into the two mass eigenstates and produce the final charged lepton and $W$. The two mass eigenstates are also going to be described as wave packets. It is only when these wave packets begin to overlap with the far detector that the charged lepton at the far detector will be measured. The  corresponding amplitudes will be non-vanishing when $t_D-t_S$ is of the order of the baseline, with uncertainties of the order of the size of the far detector. These mildly localized wave packets lead to both momentum and energy uncertainty of the same order $\Delta E  \sim \Delta p  \sim \hbar/\sigma \sim \hbar/t_S \sim 10^{-7}\,\mathrm{eV}$. These uncertainties in energy and momentum are \emph{much larger} than typical values for the neutrino energy differences $\delta m^2/\oO \sim 10^{-12}\,\mathrm{eV}$.
The issue of entanglement in neutrino oscillations was discussed within the wave packet approach   in the framework of quantum field theory in ref.\cite{aksmir2}, which argued that S-matrix calculations give rise to the correct formulas of neutrino oscillations under realistic experimental setups. Indeed, the localization of wave packets leads to results somewhat equivalent to finite time calculations, and may yield standard formulas of neutrino oscillations for most realistic experimental setups. Both the spatial and time localization of the wave packets introduced in this reference are tantamount to a finite time description, although, as mentioned above the intermediate stage invoking the S-matrix actually refers to an in-out matrix element where the initial and final time are taken to $-\infty,\infty$ respectively.  However, the interplay between   the disentangling time and the final detection time and their influence on neutrino oscillation formulas are \emph{not directly captured} by a wave packet analysis and, as we pointed out are fundamentally important and merit a thorough understanding.  \\

       The results discussed above led us to study the GSI anomaly by directly
       obtaining the decay rate of a parent and production rate of a
       daughter particle, bypassing the issue of whether amplitudes
       or probabilities for mass eigenstates must be summed. We show
       that these rates \emph{do not feature} oscillations arising
       from the interference of mass eigenstates in the final state.
       We provide an alternative field theoretical explanation in
       terms of the imaginary part of the self-energy diagram of the
       parent particle. \\

In this article, we have introduced the two time measurement approach by considering the simplest model for neutrino oscillation. It would be interesting to apply this method to the more sophisticated models. A particularly interesting model is the $2+\tilde{1}$ model of \cite{Boyanovsky:2009mq}, which features an unparticle sterile neutrino, the \emph{unsterile} neutrino, along with the two active ones. The unparticle nature of the unsterile neutrino results in four momentum dependent mixing angles and non-trivial spectral densities of the neutrino mass eigenstates. Therefore, the quantum mechanical approach to the model of \cite{Boyanovsky:2009mq} can only be thought of as a proxy description and to understand its oscillation dynamics, the quantum field theoretical two time measurement approach is necessary.



\end{document}